\documentclass[article,nofootinbib,noshowkeys,superscriptaddress]{revtex4}
\usepackage{pstricks,pst-node,pst-text,pst-3d,pslatex}
\usepackage[T1]{fontenc}
\usepackage{amsmath}

\usepackage{graphicx,amsfonts}
\usepackage{amsmath}
\newcommand{\la}{\langle}
\newcommand{\ra}{\rangle}

\newcommand{\nn}{\nonumber}

\newcommand{\LL }{\left }
\newcommand{\RR }{\right }
\newcommand{\bea}{\begin{eqnarray}}
\newcommand{\eea}{\end{eqnarray}}
\newcommand{\be}{\begin{eqnarray}}
\newcommand{\ee}{\end{eqnarray}}

\newcommand{\D}{\mathcal{D}}

\begin{document}
\author{ E. Schneider}
\affiliation{Dipartimento di Fisica, Universit\`a degli Studi di Trento, Via Sommarive 14, Povo (Trento). 38123, Italy. }
\affiliation{ Department of Chemistry, New York University, New York, New York 10003}
\author{S. a Beccara}
\affiliation{ LISC-ECT$^\star$, Via Sommarive 20 Povo (Trento). 38123, Italy.}
\affiliation{ INFN-TIFPA,  Via Sommarive 14 Povo (Trento). 38123, Italy.}
\author{F. Mascherpa}
\affiliation{Dipartimento di Fisica, Universit\`a degli Studi di Trento, Via Sommarive 14, Povo (Trento). 38123, Italy. }
\author{P. Faccioli\footnote{Corresponding author: pietro.faccioli@unitn.it}}
\affiliation{Dipartimento di Fisica, Universit\`a degli Studi di Trento, Via Sommarive 14, Povo (Trento). 38123, Italy. }
\affiliation{ INFN-TIFPA,  Via Sommarive 14 Povo (Trento). 38123, Italy.}

\title{ Quantum Propagation of Electronic Excitations in Macromolecules: A  Computationally Efficient  Multi-Scale Approach}
\begin{abstract}
We introduce a theoretical approach to study the quantum-dissipative dynamics of electronic excitations in macromolecules,  which enables to perform calculations in large systems and cover long time intervals.   All the parameters  of the underlying microscopic  Hamiltonian  are obtained from \emph{ab-initio} electronic structure calculations, ensuring chemical detail.  In the short-time regime, the theory is solvable using a diagrammatic perturbation theory, enabling analytic insight. To compute the time evolution of the density matrix at intermediate times, typically $\lesssim$~ps,  we develop a  Monte Carlo algorithm  free from any sign or phase problem, hence computationally  efficient. 
Finally, the  dynamics in the long-time and large-distance limit can be studied combining the microscopic calculations with renormalization group techniques  to define a rigorous low-resolution effective theory.  We benchmark our Monte Carlo algorithm against the results obtained in perturbation theory and using a semi-classical non-perturbative scheme. Then we apply it to compute the intra-chain charge mobility in a realistic conjugate polymer.
 \end{abstract}
\maketitle

\section{Introduction}

The problem of understanding how electronic excitations propagate through soft macromolecular systems is receiving increasing attention from physicists and chemists with diverse backgrounds, ranging from soft condensed matter physics,  to biophysics,  material science and non-linear molecular spectroscopy~\cite{charge_energy_transfer_book}. A driving motivation for this research is the perspective of designing nano-metric devices with desired optoelectronic properties, such as organic semiconductors \cite{OTFT_rev1,OTFT_rev2,organic3}, molecular wires and junctions \cite{Tao_rev,Mol_wire_review, inorganic3}. Important applications of these studies in biophysics concern the understanding of exciton energy transfer in protein photosynthetic complexes ~\cite{Engel_exp, Collini_exp, QB_Nat_rev,FMO_review} and the related perspective of designing biomimetic organic antennas. 

In all these systems,  the dynamical disorder generated by the motion of the atomic nuclei and by the  molecular environment (e.g. its solvent) 
quenches the mobility of electronic excitations, destroys their quantum coherence and introduces fluctuation/dissipation effects. 
Consequently, the electronic dynamics is not unitary and obeys the laws of open quantum systems~\cite{Petruccione_book}. 

All theoretical approaches to quantum transport and relaxation problems in macromolecules are characterized by two main ingredients: (i) the degree of resolution chosen to represent the  electronic, nuclear and environmental degrees of freedom and (ii) the algorithm used to evolve  the density matrix in time.
An incomplete list of such computational techniques includes Hierarchical Equation of Motions~\cite{Fleming_meth1,Fleming_meth2}, Time-Adaptive Density Matrix renormalization group techniques~\cite{plenio_meth1,plenio_meth2,plenio_meth3}, Non-Equilibrium Green's Function~\cite{NEGF,NEGF_book}, Path Integral Monte Carlo~\cite{Ankerhold1,Ankerhold2,Segal}, Partial Linearized Density Matrix~\cite{Coker1,Coker2} as well as methods  based on averaging the unitary evolution of the electronic excitations over an ensemble of 
independent realizations of the nuclear dynamics, obtained by means of classical molecular dynamics simulations~\cite{Aspuru_atom, Elsner, Boninsegna, Troisi_PRL, Piazza}.

Non-adiabatic \emph{ab-initio} setups  (see e.g. \cite{Agostini,Tavernelli, Garavelli})  are most theoretically rigorous and microscopically detailed, but their applicability is limited to relatively small molecules and to fast  processes. In contrast, coarse-grained models based 
on Frenkel-type tight-binding Hamiltonians or multi-scale QM-MM approaches make it possible to study larger molecular systems and to follow the time evolution even up to a few ps  \cite{Aspuru_atom, Elsner}. On the other hand, they rely on the adiabatic picture and  the electronic excitation spectrum is restricted to a single level per molecular fractional orbital.  Furthermore, additional assumptions are often introduced when coupling  electronic excitations  to the atomic nuclei and to the surrounding environment (such as  e.g. the Condon approximation). In many cases  the back-action of the electronic excitations on the vibrational dynamics is also neglected.  

Finally, phenomenological models are particularly useful for gaining qualitative insight into the physical mechanisms which shape charge transport and the related loss of quantum coherence  at different space- and time- scales \cite{Aspuru_ENAQT,plenio_FMO1,plenio_FMO2,2D_maps}. 
For example, in Ref.~\cite{Plenio_Nat} Plenio and co-workers developed a model for exciton propagation in pigment-protein complexes based on a phenomenological coupling between electronic degrees of freedom and a bath of vibrational modes. The simplicity of the model allowed them to illustrate how, in the presence of nearly degenerate exciton energy levels,  the stochastic fluctuations induced by the vibrational bath may even support electronic coherence, thus qualitatively  explaining the striking results  of non-linear optics experiments~\cite{Engel_exp,Collini_exp}. 
On the other hand, in this kind of models, the lack of a fundamental connection with the underlying microscopic chemical structure of the molecules involved makes it difficult to draw definitive conclusions about specific systems. 
 
In this work we present an attempt to combine chemical detail,  computational efficiency and physical insight in a single  ``bottom-up"  approach.
The starting Hamiltonian is basically the same adopted in Refs.~\cite{Aspuru_atom, Elsner}, based on the adiabatic separation of nuclear and electronic dynamics, on coarse-graining the electronic dynamics to the tight-binding level and on linearising the coupling between electronic excitations and molecular vibrations. All the parameters and coupling constants of this Hamiltonian can  be unambiguously obtained from \emph{ab-initio} electronic structure calculations. 

To compute the density matrix for the electronic excitations in this model we adopt a path integral representation in which the conformational dynamics of the molecule is treated in first quantization, while the electronic excitations are described using coherent fields.  In our previous papers, we developed approximations to analytically compute this path integral in the short-time limit \cite{Schneider1} (by means of  Feynman rules and perturbation theory) and in the long-time and large-distance limit~\cite{Schneider2} (using renormalization group methods). 
Here, we introduce a scheme to compute the evolution of the density matrix for the times intervals which are most relevant to compare with experiments,  i.e.  from a few fs to several ps and beyond. To this end, we introduce a Monte Carlo algorithm which exploits the fact that the conformational dynamics of a macromolecule in solution is well described in the classical limit. As a result, unlike other path integral Monte Carlo approaches to quantum transport and relaxation dynamics ~\cite{Ankerhold1,Ankerhold2}, our algorithm is not limited by any sign or phase problem, hence it is very computationally efficient.  
This  algorithm can be used to compute the effective parameters appearing in a rigorous low-resolution effective theory, obtained directly from the original Hamiltonian using  Renormalization Group (RG) techniques. The simplicity of such an effective theory makes it possible to investigate the dissipative dynamics of electronic excitations in mesoscopic systems. 

Our approach displays a few additional useful features. First, it enables to rigorously account  for the quantum back-action exerted by the electronic excitations on the dynamics of the nuclei. 
In addition, it allows us to single out the effect of the coupling of the electronic degrees of freedom with an arbitrary chosen \emph{subset} of  vibrational normal modes. We recall that, in the underlying model,  the coupling constants and the normal mode frequency spectrum are microscopically  derived directly from electronic structure calculations. Hence,  our method may be applied to test from first principles if the  propagation of the electronic excitations and the quantum decoherence/recoherence processes in a given molecule are strongly affected by specific  vibrational modes. 
We emphasize that, even though  intermediate derivations  are based on the quantum field theory and path integral formalism, the final results and the numerical algorithm are defined in terms of standard quantum mechanical objects and can be straightforwardly implemented.

We first  benchmark our different approaches by studying a model for exciton relaxation dynamics in a linear molecular chain which was recently proposed  and solved by Iubini \emph{et. al.} using an alternative semi-classical approximation \cite{Piazza}. Next, we present a realistic study of intra-chain  hole propagation  in a  long isolated conjugated polymer in solution. In this calculation, all effective parameters are derived directly from quantum electronic structure, while the matching to the low-resolution approximation enables to explore charge propagation at  mesoscopic scales. 

The paper is organized as follows. In section \ref{model} we introduce the microscopic model Hamiltonian. In sections \ref{PIderivation} and \ref{relativisticnotation} we review the derivation of the field theoretic expression for the reduced density matrix, and we compute the effective action describing the effect of electronic excitations on the vibronic dynamics. In section \ref{perturbative} we discuss the perturbative approximation, which enables to obtain analytically the density matrix at weak coupling  or at short times. 
In section, \ref{non-perturbative} we introduce our non-perturbative Monte Carlo approach. 
Readers not interested in technical field-theoretic derivations can find a self-contained summary and discussion of the main equations and of the relative Monte Carlo algorithm at the end of this section, which only makes use of standard quantum mechanical concepts. In section \ref{ETtheory} we discuss the matching of microscopic calculations with a rigorous low-resolution effective theory. 
Illustrative applications are presented and discussed in sections \ref{application1} and  \ref{application2} and the main results and conclusions are summarized in section \ref{conclusions}.

\section{A Microscopic Model for Quantum Transport in Open Macromolecular Systems}\label{model}

Throughout most of this work we shall rely on a standard microscopic model for quantum transport in macromolecules, based on coarse-graining the electronic dynamics at the level of molecular orbitals. We choose to introduce this model using the language of RG and  Effective  Theory (ET) formalism. This  allows us to set the stage for the multi-scale approach which will be introduced in section \ref{ETtheory}. 

Effective theories  ---for an excellent introduction see \cite{Lepage}--- are systematic low-energy approximations of more microscopic theories. They are based on the familiar observation that an experimental probe with wavelength $\lambda$ is insensitive to any physical details at the so-called Ultra-Violet (UV) scales, i.e.  $\ll \lambda$. 
This observation can be used to construct approximations of the original microscopic dynamics which hold in the opposite Infra-Red (IR) limit, i.e. at scales $\gg \lambda$, where $\lambda$ plays the role of a  cutoff.
In such low-energy approximations, all the  physics above the UV cutoff is mimicked by  a set of local operators and effective parameters. Indeed, at the resolution set by the cutoff  $\lambda$, the difference between the correct (i.e. microscopically detailed)  description of the UV dynamics and the approximate one (defined by the effective operators) is not detectable. 

Clearly, the ET approach can only hold if the UV and IR  scales are decoupled. In quantum systems, this condition is typically realized when the spectrum of the underlying microscopic Hamiltonian is gapped. In this case, ETs provide approximations for the low-energy dynamics of the quantum states below the gap. 

 In the specific case of quantum dissipative dynamics in macromolecules, we assume the existence of two characteristic UV scales: $\sigma_{\textrm short}$ and $\tau_{\textrm short}$.   $\tau_{\textrm short}$ is the typical time  it takes an electronic excitations takes to delocalize over molecular fractions of size $\sigma_{\textrm short}$ much smaller than the molecular size $S$. Denoting with $t$ the typical IR  time scale  at which electronic excitations spread over the entire molecule, we assume that $\tau \gg \tau_{\textrm short}$.   We want to develop an ET with two UV cutoffs:  $\lambda\gtrsim \sigma_{\textrm short}$ (typically in the nm scale) and $t \gtrsim \tau_{\textrm short}$ (typically in the fs scale). Then, the quantum electronic state after the fast relaxation within each of the $N$ fragments  of size $\sigma_{\textrm short}$ can be described by a set of molecular fragment orbitals $|\Phi_1\ra, \ldots |\Phi_N\ra$.

The decoupling between $\tau_{\textrm{short}}$ and $t$ is realized if the propagation of an electronic excitation over the entire molecule occurs through quantum  tunnelling across the molecular fragments. Then the spectrum of the total electronic Hamiltonian is gapped, and the number of energy levels below the gap coincides with the  number  of  molecular fragments. Then the low-energy quantum dynamics associated with the states below the gap can be well described by diagonalising the original microscopic  Hamiltonian in the model space of molecular fragment orbitals, $|\Phi_1\ra, |\Phi_2\ra, \ldots$

At the time-resolution scale $\tau_{\textrm{short}}$, it is legitimate to adopt the adiabatic approximation. 
In addition, at the  space-resolution scale $\sigma_{\textrm short}$, the dynamics of the individual atoms is not resolved. Instead, the  conformational dynamics  is described by the set of generalized coordinates $Q\equiv (q_1, q_2,\ldots )$, which specify  the molecular conformation with a space resolution $\gtrsim \sigma_{\text{short}}$. For example,  they  may be the Cartesian coordinates of beads which collectively represent groups of atoms,  or may be a set of  relevant dihedral angles. 
 
Having specified the resolution power of our theory, we are finally in a condition to introduce our model  Hamiltonian, which consists of the following three terms:
\be\label{Htot}
\hat H = \hat H_{MM}+  \hat H_{CL}+ \hat H_{FH}.
\ee
The first term describes the molecule's conformational dynamics and reads (using Cartesian coordinates for sake of definiteness): 
\be\label{HMM}
\hat H_{MM} \equiv \sum_{i=1}^{3 N}\frac{\hat p_i^2}{2 M} + \hat V(q_1, \ldots, q_{3 N} ) \, .
\ee
In this equation, $N$ is the number of beads and  we have taken all masses to be the same for sake of notational simplicity. $V(Q)$ is the molecular potential energy function which parametrizes the lowest Born--Oppenheimer energy surface. Note that,  in the classical approximation, the Hamiltonian (\ref{HMM}) gives rise to the standard classical molecular mechanics for an isolated molecule in the gas phase.
 
The second term in Eq. (\ref{Htot}) describes the coupling between the set of coordinates  $Q$ and the environment (i.e. the solvent).  As we shall see below, at low time resolutions, the specific form of $H_{CL}$ is not crucial. For sake of convenience, we define $H_{CL}$ to be the standard  Caldeira--Leggett Hamiltonian, which models the environment as an infinite set  of harmonic oscillators: 
 \be
 \hat H_{CL}=\sum_{i=1}^{3N} \sum_{\alpha=1}^\infty \left(\frac{\hat \pi^2_\alpha}{2 \mu_\alpha} + \frac{1}{2} \mu_\alpha \omega_\alpha^2 \hat x_\alpha^2 - c_\alpha  \hat x_\alpha \hat q_i +
 \frac{c_\alpha^2}{2 \mu_\alpha \omega_\alpha^2}~\hat q_\alpha^2\right) \, .
 \ee
The Hamiltonian $\hat H_L\equiv \hat H_{MM}+  \hat H_{CL}$ is defined in such a way to generate the standard Langevin dynamics for the set of generalized coordinates $Q$ in the so-called Ohmic limit for the spectral density of the harmonic oscillators and in the classical limit for $Q$. 

 The third term is the Frenkel Hamiltonian which correctly describes the dynamics of the electronic excitations at our level of space and time resolution:
\be\label{HFH}
 \hat H_{FH} = \sum_{{ m}, { n}} ~\la \Phi_n| \hat H_{el} |\Phi_{ m} \ra ~\hat a_{ n}^\dagger \hat a_{m } \, , 
\ee  
where $|\Phi_{ m}\ra$ are the fractional molecular orbitals labeled by the discrete index $m$ and $\hat H_{el}$ is the Coulomb Hamiltonian. 
In the adiabatic limit, the matrix elements $ \la \Phi_n| \hat H_{el} |\Phi_{ m} \ra $ depend parametrically on the generalized coordinates $Q$, i.e. 
\be
 \la \Phi_n| \hat H_{el} |\Phi_{ m} \ra \equiv  f_{n m}(Q).
\ee
We note that, in order  to account for Landau--Zener type transitions,  the Frenkel Hamiltonian should be extended to include terms proportional to a single creation or annihilation operator.  However, in this work we choose to ignore such terms since we intend to focus on molecular systems  in which the electronic excitation dynamics are long-lived and confined on a single Born--Oppenheimer surface.
 
The total Hamiltonian $\hat H$ describes the complete dynamics of the macromolecule in its heat bath. 
However, we are focused on studying the time evolution of the electronic excitations,  hence both the heat bath variables and the generalized coordinates $Q$ can be traced out from the density matrix.  
The dynamics of the electronic excitation is then entirely encoded in the so-called reduced density matrix:
\be
\rho^R_{ nm}(t) \equiv 
 \sum_{Q} \sum_{X}%
 \la n, Q, X |~\hat \rho(t) | m, Q, X \ra \equiv \text{Tr}_{X,Q}[|m \ra \la n |\hat \rho(t)] \, ,
\ee
In the next section we review a convenient path integral representation for $ \rho^R_{nm}(t) $, which was originally derived in our previous work~\cite{Boninsegna, Schneider1} and provides the suitable framework to develop our Path Integral Monte Carlo approach. For all details and derivations which are not reported here we refer the reader to the original publication~\cite{Schneider1}.

\section{Coherent Field Path Integral Representation of the Reduced Density Matrix}\label{PIderivation}
The starting point consists in explicitly writing the system's time evolution operators and the initial condition
 \be\label{time}
 \rho^R_{n m}(t) = \text{Tr}_{Q,X}[|m \ra \la n |e^{-\frac{i}{\hbar} \hat H t} ~\hat \rho(0)~e^{\frac{i}{\hbar} \hat H t} ] \, .
 \ee
In particular, in the following we shall consider initial density matrixes in the form
 \be\label{imtime}
\hat  \rho(0) =  |n_0\ra \la n_0| \times e^{-\frac{1}{k_B T} \left(\hat H_{MM}+\hat H_{CL}\right)}.
 \ee
 This corresponds to assuming that one electronic excitation is created a time $t=0$, while the macromolecule was in the electronic ground state and in equilibrium with the heat bath. 
 
The path integral for the density matrix can be obtained through the standard Trotter decomposition technique applied to the two real-time evolution operators in Eq.~(\ref{time}) and to the imaginary time evolution operator representing the initial thermal density matrix in Eq. (\ref{imtime}). 
To this end we need to insert a representation of  the identity in terms of the system's quantum states  at each elementary step in the real- and imaginary time evolution. We choose the following form \cite{Negele}:
\be \label{Trotter}
1  = \int d Q \int d X \int \prod_{l} \frac{d\phi_l d\phi^*_l}{2\pi i} |Q,X,\phi_l\ra \la Q,X,\phi_l | e^{-\sum_k \phi_k^* \phi_k} \, ,
\ee
where $  |\phi_k  \ra$ are the coherent states associated with the electronic excitation,  $l$ is the index  labelling the fractional orbital and the exponential factor accounts for the over-completeness of this basis. $|Q\ra$  and $|X\ra$ denote the position eigenstates of the beads (assuming for simplicity a Cartesian representation for the conformational dynamics)  and of the set of Caldeira-- Leggett variables, respectively.  
We shall always assume that molecular fractional orbitals host at most a single electronic excitation. In this case,  spin-statistics correlations are irrelevant and we can adopt the bosonic coherent field description. 

The path integral representation of the density matrix is obtained as usual by inserting the identity at each elementary evolution and then taking the continuum limit. 
After performing the Gaussian integration over the Caldeira--Leggett variables the  density matrix takes the form:
\be\label{PI0}
\rho^R_{n m} (t) &=&\int  \mathcal{D} Q' \mathcal{D} Q'' \mathcal{D} \bar Q \mathcal{D} \phi' \mathcal{D} \phi'' ~
 \mathcal{D} \phi^{'*}\mathcal{D} \phi^{''*}
~\left(\phi'_n(t) \phi^{'*}_{n_0}(0) \phi^{''*}_{m}(t) \phi^{''}_{m_0}(0)\right)~ e^{-\sum_m (\phi_m^{''*}(t) \phi^{''}_m(t)+ \phi^{'*}_m(0)\phi^{'}_m(0) )} \nn\\
&& e^{\frac{i}{\hbar}~\left(S_{MC}[Q^{'},\phi^{'},\phi^{'*}]- S_{MC}[Q^{''},\phi^{''},\phi^{''*}]\right)}\times ~e^{\frac{i}{\hbar}\Phi[Q',Q'']}\times e^{- \frac{1}{\hbar} S_{E}[\bar Q]} \
\ee
In this expression, the superscripts $'$ and $''$ refer respectively to the degrees of freedom associated with the upper and lower branches of the Keldysh contour
i.e. to coherent states and nuclear positions evolved by  $e^{-\frac{i}{\hbar} H t}$ and  $e^{\frac{i}{\hbar} H t}$ respectively. 
Instead,  $\bar Q$ propagates along the imaginary time axis, enters in the representation of the initial thermal density matrix. 
The prefactor in round brackets arises from the fact that the initial and final conditions contain one electronic excitation. The functionals which appear at the exponent of Eq. (\ref{PI0}) are defined as 
\be\label{actions}
S_{E}[\tilde Q] &=& \int_0^{\beta \hbar} d\tau \left( \frac{M}{2} \dot{\tilde{Q}}^2(\tau)+ V[\tilde Q(\tau)] \right), \\
S_{MC}\left[Q, \phi, \phi^{*}\right]  &=&  \int_{T}^{T+t} dt'~\left\{ \frac{M}{2} \dot Q^2(t')- V[Q(t')]
+  \sum_{mn} \phi_{m}^*(t')\left( i \hbar \frac{\partial }{\partial t'}\delta_{m n} -f_{m n}[Q(t')] \right) \phi_{n}(t')\right\} \, , \\
\Phi[Q',Q'']&=& i \int_{T}^{T+t} dt' \int_0^{t'} dt'' \left\{ \left(Q'(t')-Q''(t') \right)\cdot\left[\mathcal{B}(t'-t'')Q'(t'') - \mathcal{B}^*(t'-t'')Q''(t'') \right]\right\}\nonumber\\
&+&  i\frac{\bar\mu}{2}\int_{T}^{T+t} dt'\left[{Q'}^2(t')-{Q''}^2(t') \right], \qquad  \left(\bar\mu = \frac{c_j^2}{m_j \omega_j^2}\right) \, . 
\ee
The two-point function $\mathcal{B}(t-t')$ appears upon integrating out the heat bath variables and reads
\be\label{B}
\mathcal{B}(t-t') = \sum_\alpha \frac{c^2_\alpha}{\mu_\alpha \omega_\alpha}\left[\text{coth}\left( \frac{\omega_\alpha}{2 k_B T}\right) \cos(\omega_\alpha t)- i \sin(\omega_\alpha t) \right].
\ee
This Green's function encodes the characteristic response of the heat bath (e.g. the water surrounding the molecule) to the molecular motion.

In order to make this expression numerically tractable, we apply the following  series of well defined approximations: 
\begin{enumerate}
\item \emph{Classical limit for the conformational dynamics}: 

 It has long been established  that in a large class of relevant molecular processes  the conformational dynamics is accurately described by the classical approximation, i.e. through the so-called  molecular dynamics. Notable counter-examples are transitions involving  quantum tunnelling of the nuclei,  such as proton transfer reactions.   In the following, we do not consider such class of problems and exploit the simplifications which emerge from treating the effective degrees of freedom $Q$ at the classical level.

The path integral formalism provides a very convenient framework to implement the classical approximation on the dynamics of the generalized coordinates $Q$, while retaining a full quantum treatment of the electronic  dynamics. 
This is done by rewriting  the path integral (\ref{PI0}) in a form in which forward and backward 
molecular paths $Q'(t)$ and $Q''(t)$ are replaced by their average and difference:
\be
R(t)\equiv \frac{Q'(t)+Q''(t)}{2} \, ,\qquad y(t) \equiv Q''(t)-Q'(t),
\ee
and observing that $|y(t)|\ll |R(t)|$ throughout  the regime of masses and temperatures in which  the classical approximation is justified.  The classical approximation is then obtained  by expanding the action functionals (\ref{actions}) up to quadratic order in $y(t)$ and then performing the resulting Gaussian path integral in $\mathcal{D}y$ in Eq. (\ref{PI0}). The result is an effective theory  in which the nuclear dynamics is described only by the integral over the $R(\tau)$ variables, which are interpreted as the classical nuclear configurations sampled in a generalized Langevin dynamics with noise memory kernel $B(t-t')$. 

\item \emph{Ohmic bath limit for the solvent:} 
We further  simplify  our theory by imposing that the conformational dynamics is described by a  Langevin equation with delta-correlated white noise, as is routinely done in  MM simulations. To this end we consider a slow-frequency expansion of the  Fourier transform of the Caldera-Leggett two-point function $B(t-t')$:
\be\label{ohmic}
\tilde B(\omega) =  B_0 + B_1 \omega + \ldots,\ee 
which, in time representation, corresponds to $\mathcal{B}(t)\simeq  B_0 \delta(t)+ i B_1 \frac{d}{dt} \delta (t).$
The coefficients $B_0$ and $B_1$  are determined by imposing that, in the classical limit, the probability of observing a particle some specific position $R$ at time $t$, $P(R,t)\equiv \la R| \hat \rho(t)| R\ra $ must obey a Fokker-Planck equation with friction coefficient $\gamma$ and temperature $T$. The result is
\be
B_0 = 2 k_B T M \gamma, \qquad B_1 = M\gamma.
\ee

\item \emph{Small oscillations}: In the present work we are primarily interested in the quantum dynamics of the electronic excitations. In most such cases, within   the characteristic  timescales for electronic energy transfer processes, macromolecules  cannot perform large structural  rearrangements, but only undergo small thermal oscillations  around their equilibrium configuration. 
Hence, we can Taylor-expand  the   positions to linear order near the mechanical equilibrium configuration $R_0$
\be
R_i = R_i^0 + \delta r_i(t) \, .
\ee
Consistently, we expand the molecular potential energy to  quadratic order around $R_0$: 
\be
V(R) = V(R_0) + \frac{1}{2} \mathcal{H}_{ij} \delta r_i \delta r_j \qquad \left(\mathcal{H}_{ij} \equiv \lim_{R\to R_0} \frac{\partial}{\partial R_i} \frac{\partial}{\partial R_j} V(R)\right)
\ee
and  linearize also the overlap matrix elements:
\be
f_{nm}(R) = f^0_{nm} + \sum_i f^i_{n m}~\delta r_i \, ,
\ee
where 
$
f^0_{nm}\equiv f_{nm}(R_0)$ and $f^i_{n m} \equiv  \frac{\partial}{\partial R^i} f_{nm}(R_0)$.

 Applying this set of approximations to the original path integral (\ref{PI0}) leads to the following expression for the density matrix (for a detailed derivation we refer the reader to Ref. \cite{Schneider1}):
\be\label{PI1}
&&\rho_{n m}(t) \simeq \textrm{const}\times \int \mathcal{D} \phi' \mathcal{D} \phi^{'*} \mathcal{D} \phi'' \mathcal{D} \phi^{''*} \int \mathcal{D} \delta r_i~\cdot~e^{-\beta H_{MM}(0)}~\left(\phi'_n(t) \phi^{'*}_{n_0}(0) \phi^{''*}_{m}(t) \phi^{''}_{m_0}(0)\right)~e^{-\sum_m (\phi_m^{''*}(t) \phi^{'}_m(t)+ \phi^{'*}_m(0)\phi^{'}_m(0) )} \nonumber\\
&&\cdot e^{\frac{i}{\hbar}~ \sum_{mn}  \int_0^t dt'   
\left[ i~\hbar\frac{\partial}{\partial t'}\delta_{n m} - f_{nm}^0- \sum_i \delta r_i f^i_{nm} \right] 
(\phi_{n}^{\ast '}(t') \phi^{'}_{m}(t')-\phi_{n}^{\ast ''}(t') \phi^{''}_{m}(t'))}
 \cdot e^{-\frac{\beta}{4 M \gamma} \int_{0}^{t} d\tau\sum_i \left\{M \delta\ddot{ r}_i + M \gamma \delta \dot{ r}_i + \sum_j \delta r_j \mathcal{H}_{ij}  + \sum_{nm} \frac{f^i_{nm}}{2}(\phi^{'*}_n\phi'_m + \phi^{''*}_n\phi''_m) \right\}^2},\nn\\
\ee
 where 
 \be
 H_{MM}(0)) \equiv  \frac{1}{2}\sum_{ij} \mathcal{H}_{ij}~\delta r_i(0) \delta r_j(0) + \frac{M}{2}\sum_{i}~\delta \dot r^2_i(0)
 \ee 
  is the value of the classical  Hamiltonian for the molecular conformation at $t=0$.
We note that if the coupling between vibrational and electronic excitations is neglected, the path integral factorizes into a vibronic and excitonic part. The integral in the vibronic coordinates reduces to the well-known Onsager--Machlup (OM) path integral representation of the classical Langevin dynamics.


Let us now analyse the couplings between the electronic and vibronic excitations. After expanding the square in the OM functional the density matrix is rewritten as
\be\label{PI2}
\rho_{n m}(t) &\simeq& \int \mathcal{D} \phi' \mathcal{D} \phi^{'*} \mathcal{D} \phi'' \mathcal{D} \phi^{''*} \int \mathcal{D} \delta r_i~\left(\phi'_n(t) \phi^{'*}_{n_0}(0) \phi^{''*}_{m}(t) \phi^{''}_{m_0}(0)\right)~e^{-\sum_m (\phi_m^{''*}(t) \phi^{'}_m(t)+ \phi^{'*}_m(0)\phi^{'}_m(0) )} \nonumber\\
&\cdot& e^{\frac{i}{\hbar}~ \sum_{mn}  \int_0^t dt'   
\left[ i~\hbar\frac{\partial}{\partial t'}\delta_{n m} - f_{nm}^0 \right] 
(\phi_{n}^{\ast '}(t') \phi^{'}_{m}(t')-\phi_{n}^{\ast ''}(t') \phi^{''}_{m}(t'))}\cdot 
e^{-\frac{\beta}{4 M \gamma} \int_{0}^{t} d\tau\sum_i \left\{M \delta\ddot{ r}_i + M \gamma \delta \dot{ r}_i + \sum_j \delta r_j \mathcal{H}_{ij}  \right\}^2}
\nn\\
&\cdot& e^{\frac{i}{\hbar} S_{int}[\delta r, \phi',\phi^{'\ast}, \phi'', \phi^{''\ast}]} \cdot e^{-\beta H_{MM}(0)}
\ee
where the interaction action is
\be
S_{int}[\delta r, \phi',\phi^{'\ast}, \phi'', \phi^{''\ast}] &=&\int_{0}^{t} d\tau \left\{  \frac{i \hbar \beta}{4 M \gamma} \left[
\left(\sum_i J_i \right)^2+2 \sum_i \left(M \delta\ddot{ r}_i + M \gamma \delta \dot{ r}_i + \sum_j \delta r_j \mathcal{H}_{ij}\right) J_i
 \right.\right] \nn\\
&&\left.-\sum_{nm}\sum_i\delta r^i f_{nm}^i (\phi^{'*}_n\phi'_m - \phi^{''*}_n\phi''_m) \right\}.
\ee
where $J_i \equiv \sum_{nm}\frac{f^i_{nm}}{2}(\phi^{'*}_n\phi'_m + \phi^{''*}_n\phi''_m)$.
We note that the first line is proportional to the factor $\propto \frac{\hbar \beta}{M}$ which suppresses these interactions in the regime of temperature and  masses in which the classical approximation to the conformational dynamics  is applicable. Therefore, we can consistently drop these couplings and consider only the third term
\be
S_{int}[\delta r, \phi',\phi^{'\ast}, \phi'', \phi^{''\ast}] &\simeq&- \int_{0}^{t} d\tau\sum_i \sum_{nm}\delta r^i f_{nm}^i (\phi^{'*}_n\phi'_m - \phi^{''*}_n\phi''_m).
\ee


\end{enumerate}

\section{Formal Analogy with Relativistic Quantum  Field Theory}
\label{relativisticnotation}
 
The field degrees of freedom $\phi''$, which are associated with the lower branch of the  Keldysh contour, describe excitons  evolving according to the $e^{\frac{i}{\hbar} H t}$ operator, i.e. propagating  ``backward" in time. This observation enables us to establish a \emph{formal} analogy between the non-unitary evolution of a single exciton propagating through a macromolecules, with the dynamics of a relativistic particle (with both matter and anti-matter components) propagating in vacuum. 
More precisely, the path integral (\ref{PI2})  is formally equivalent to the expression of the time-ordered two-point Green's function of a relativistic particle-antiparticle system in vacuum \cite{Schneider1} . 

To make this formal analogy completely explicit, it is sufficient to  organize the coherent fields into a doublet:
\be
\psi^a_n(t) = \left(\begin{array}{c}
\phi'_n(t)\\  \phi''_n(t)
\end{array} \right)\, ,
\ee
where the index $a=1,2$ labels the upper and lower components and its summation will be implicitly assumed whenever not displayed.  The two components of the doublet correspond to the two branches of the Keldysh contour, and  they will be respectively identified   as the fictitious matter and anti-matter components in the corresponding virtual relativistic quantum field theory. 

Let us also  introduce the following  matrices 
\be
 \gamma_0 = \left( \begin{array}{cc} 1 &0 \\ 0 & -1\end{array}\right) \quad \gamma_5 =\left( \begin{array}{cc} 0 &1 \\ 1 & 0\end{array}\right)\quad \gamma_{+}  = \left( \begin{array}{cc} 1 &0 \\ 0 & 0\end{array}\right)\quad  \gamma_{-} =\left( \begin{array}{cc} 0 &0 \\ 0 & 1\end{array}\right)\, .
\ee
and adopt the Dirac notation $\bar \psi = \psi^\dagger \gamma_0$.  Then, the  path integral for the one-excitation density matrix  (\ref{PI2}) becomes
\be\label{rhoQFT}
 \rho_{nm}(t)=\int \D\delta r  \int \mathcal{D}\bar \psi \mathcal{D} \psi~O^{+}_{nm}(t)~O^{-}_{n_0m_0} (0)~e^{-S_{OM}[\delta r]}~e^{\frac{i}{\hbar} \left(S_0[ \psi, \bar \psi] +  S_{int}[\psi, \bar \psi; \delta r] \right) + \mathcal{L}} \,~e^{-\beta H_{MM}(0)},
\ee
where $\mathcal{L}\equiv \sum_m \left( \bar \psi_m(0) \gamma_0 \gamma_+ \psi_m(0)+  \bar \psi_m(t) \gamma_0 \gamma_- \psi_m(t)\right)$ is the standard surface term due  to the  over completeness of the coherent state basis and the functionals at the exponent are defined as follows:
\be
\label{SOM}
S_{OM}[\delta r]&\equiv& \frac{\beta}{4 M \gamma}\sum_i \int_{0}^{t} d\tau~\left(M \delta\ddot{ r}_i + M \gamma ~\delta \dot{ r}_i + \sum_j \delta r_j \mathcal{H}_{ij}  \right)^2\\
\label{S0}
 S_0[ \psi, \bar \psi]  &=& \sum_{nm} \int_{0}^{t} d\tau ~ \bar \psi_{n}(\tau) ~\left(i \hbar  \delta_{nm} \partial_t - f_{n m}^0\right) ~\psi_m(\tau)\, ,\\
\label{iphi}
S_{int}[ \psi, \bar \psi] &=& - \sum_{nm} \sum_{i} \int_{0}^{t}d\tau~\delta r_i(\tau)~ \bar \psi_{n}(\tau) f^i_{nm} \psi_{m}(\tau).\ee
Finally, the bilinear ``densities" in front of the exponent read
\be\label{O+-}
 O^{+/-}_{nm} (\tau)= \bar \psi_n(\tau) \gamma_5 \gamma_{+/-} \psi_m(\tau)\, .
\ee



In order to obtain the expression of density matrices containing  an arbitrary number of electronic excitations it is  convenient to introduce a generating functional:
\be\label{generating}
Z[\eta, \bar \eta] = \int \D\delta r ~e^{-S_{OM}[\delta r]} \int \mathcal{D}\bar \psi \mathcal{D} \psi~e^{\frac{i}{\hbar} \left(S_0[ \psi, \bar \psi]+S_{int}[\psi, \bar \psi;\delta r]\right) + \mathcal{L} }~
e^{ \frac{i}{\hbar}\sum_n\int_{0}^{t} d\tau (\bar \eta_{n}(\tau) \psi_{n}(\tau) + \bar \psi_{n}(\tau) \eta_{n}(\tau))}~ ~e^{-\beta H_{MM}(0)}
\ee
where $\eta_n(\tau)$ and $\bar \eta_n(\tau)$ are external source fields. 
The expression for arbitrary density martices with $n$  excitons are  obtained by appropriate combinations of functional derivatives with respect to the external sources. For example, the single excitation density matrix (\ref{rhoQFT}) is obtained as follows:
\be\label{rho}
\rho_{n m}(t)  = \lim_{\eta, \bar \eta\to0} \frac{(\gamma_5 \gamma_-)_{dc}~(\gamma_5 \gamma_+)_{ba}}{Z[\bar \eta, \eta]} \hbar^4 \frac{ \delta}{\delta \bar \eta^a_m(t)}  \frac{ \delta}{\delta \eta^b_{n}(t)}\frac{ \delta}{\delta \bar \eta^c_{n_0}( 0)}  \frac{ \delta}{\delta \eta^d_{m_0}( 0)} Z[\bar \eta, \eta] \, .
\ee

 Finally, we note that the path integral over the vibronic variables $\delta r_i(\tau)$ is Gaussian and can be carried out analytically. The result is 
an equivalent effective theory formulated in terms of exciton degrees of freedom only, described by the generating functional 
\be\label{generating2}
Z[\eta, \bar \eta] = \int \mathcal{D} \psi \mathcal{D} \psi~e^{\frac{i}{\hbar} S_0[ \psi, \bar \psi]+ \mathcal{L}}~
e^{-\frac{M~\gamma}{\beta \hbar^2}~\int_0^t d\tau \int_0^t d\tau' \sum_{n m k l}\sum_{ij} ~f^i_{nm} ~f^j_{k l}~\bar \psi_n(\tau) \psi_m(\tau)~\Delta_{ij}(\tau,\tau')~\bar \psi_k(\tau') \psi_l(\tau')} ~e^{ \frac{i}{\hbar}\sum_n\int_{0}^{t} d\tau (\bar \eta_{n}(\tau) \psi_{n}(\tau) + \bar \psi_{n}(\tau) \eta_{n}(\tau))}~
\ee
where $\Delta_{ij}(\tau, \tau')$ is the correlation function of vibronic excitations propagating according to the Langevin dynamics:
\be
\Delta_{ij}(\tau, \tau') &=& \langle \delta r_i(\tau) \delta r_j(\tau') \rangle \nn\\
&=& \frac{\int d\delta r_0  e^{-\beta \frac{1}{2}~\sum_{ij} ~\delta r^i_0 \mathcal{H}_{ij} \delta r^j_0}
\int_{\delta r_0} \mathcal{D} \delta r~ (~\delta r_i(\tau)~ \delta r_j(\tau')~)~e^{-\frac{\beta}{4 M \gamma} \int_{0}^{t} d\tau\sum_i \left\{M \delta\ddot{ r}_i + M \gamma \delta \dot{ r}_i + \sum_j \delta r_j \mathcal{H}_{ij}  \right\}^2} }{\int d\delta r_0  e^{-\beta \frac{1}{2}~\sum_{ij} ~\delta r^i_0 \mathcal{H}_{ij} \delta r^j_0}
\int_{\delta r_0} \mathcal{D} \delta r~e^{-\frac{\beta}{4 M \gamma} \int_{0}^{t} d\tau\sum_i \left\{M \delta\ddot{ r}_i + M \gamma \delta \dot{ r}_i + \sum_j \delta r_j \mathcal{H}_{ij}  \right\}^2}}.
\ee
This correlation function can be explicitly evaluated:
\be\label{Delta}
\Delta_{ij}(\tau, \tau')
&=& \frac{e^{-\frac{1}{2}\gamma~|\tau-\tau'|}}{2~M^2}~\sum_k~ U_{i k}~U^\dag_{k j}~\frac{a_{k}(\tau-\tau')}{\Omega^2_k} \, ,\\
\ee
where $U_{jk}$ is the orthogonal matrix which diagonalizes the Hessian matrix $\mathcal{H}_{ij}$, $\Omega_k$ are the corresponding normal mode frequencies and $\omega_0^k=\sqrt{\LL|4 \Omega_k^2 - \gamma^2 \RR|}$. Finally, the functions $a_k(\tau-\tau')$ are defined by:
\be\label{ak}
a_k(t)&=&\left\{ 
\begin{array}{cc}
 \displaystyle{ \frac{\sin  \LL( \frac{1}{2} \omega_0^k~|t| \RR)}{\omega_0^k}  +  \frac{\cos  \LL( \frac{1}{2} \omega_0^k~|t| \RR)}{\gamma} }  & \quad  \text{if} ~ 2 \Omega_k \geq \gamma \,,  \\ 
 \displaystyle{ \frac{\sinh \LL( \frac{1}{2} \omega_0^k~|t| \RR)}{\omega_0^k}  +  \frac{\cosh \LL( \frac{1}{2} \omega_0^k~|t| \RR)}{\gamma} }  & \quad \text{if} ~ 2 \Omega_k < \gamma \, , 
\end{array}\right.
\ee
We emphasise that the summation over the vibrational zero-modes $\Omega_k$ in Eq. (\ref{Delta}) should not include the zero-mode $\Omega_0=0$, which is associated with center of mass diffusion.  

The dissipative interactions  provided by the functional
\be
S_{int} [\bar \psi, \psi] \equiv -\frac{M~\gamma}{\beta \hbar^2}~\int_0^t d\tau \int_0^t d\tau' \sum_{n m k l}\sum_{ij} ~f^i_{nm} ~f^j_{k l}~\bar \psi_n(\tau) \psi_m(\tau)~\Delta_{ij}(\tau,\tau')~\bar \psi_k(\tau') \psi_l(\tau') 
\ee
are responsible for translating vibronic correlations into electronic correlations. In particular, fast underdamped  vibronic normal modes  ---see first line of Eq. (\ref{ak})---  can induce coherent oscillations in the electronic density matrix. This effect will be clearly visible in the illustrative examples  discussed in sections \ref{application1} and \ref{application2}.

\section{Perturbative calculation of the density matrix}\label{perturbative}

\begin{figure}[t!]
\includegraphics[width=12 cm]{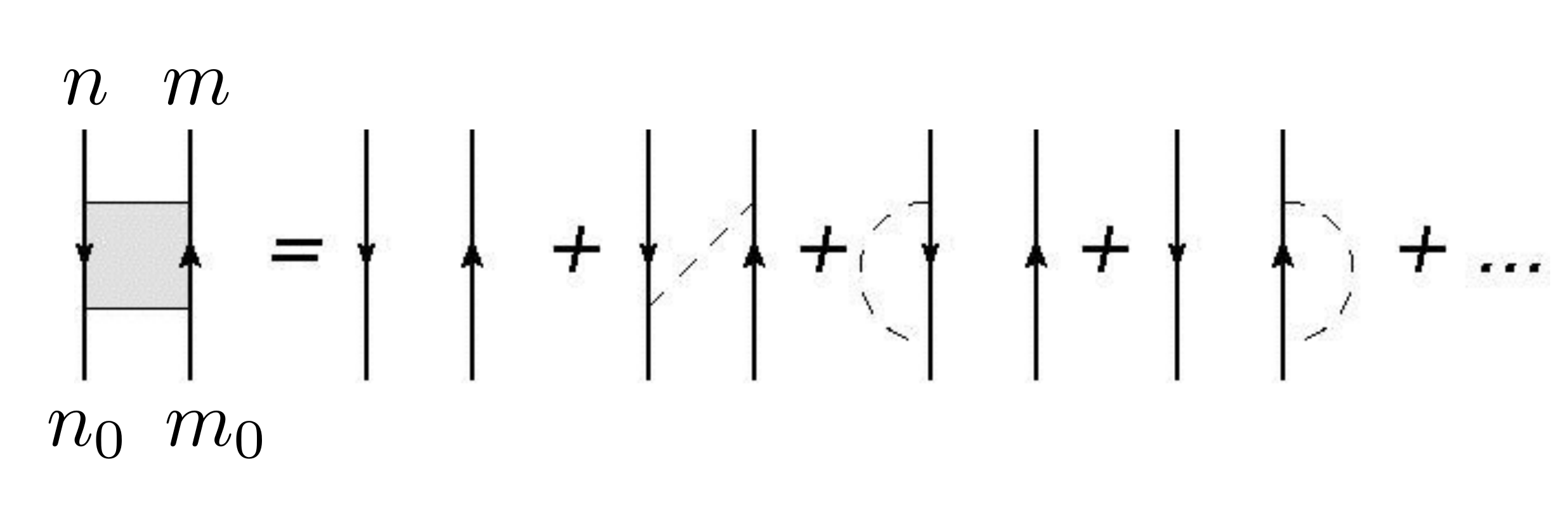}
\caption{Diagrammatic representation of the density matrix for freely evolving electronic excitations. The density matrix is formally equivalent to a two-point function in the dual relativistic particle-antiparticle system.}
\label{two point}
\end{figure}

If the small coupling regime or for very short time intervals $t$, it is possible to estimate the reduced  density matrix of the excitons  in perturbation theory. 
 An useful feature of the relativistic quantum field theory notation is that it enables to compute perturbative corrections applying standard diagrammatic techniques~\cite{Schneider1}. For example, the Feynman diagrams which appear in the leading order correction to the evolution of the density matrix element $\la n| \hat \rho(t)| m\ra$, starting from the initial condition $\hat \rho(0) = |n_0\rangle \langle m_0|$  are shown in Fig.\ref{two point}. In these diagrams, the dashed line denotes the vibronic propagator $\Delta_{ij}(\tau,\tau')$ given in Eq. (\ref{Delta}).
 The solid lines represent the time ordered exciton propagator, which is defined as: 
\be
G^0_{ab}(n,\tau|m,\tau') &=&
 \frac{ \int \mathcal{D}\bar \psi \mathcal{D} \psi~\psi^a_{n}(\tau)~\bar \psi^b_{m}(\tau')~ 
  e^{\frac{i}{\hbar} \sum_{mn}  \int_{0}^{t} d\tau \bar \psi_{m}(\tau) \left[ i \hbar \partial_t\delta_{mn} - f^0_{m n} \right]\psi_{n}(\tau)} }
 { \int \mathcal{D}\bar \psi \mathcal{D} \psi~ e^{\frac{i}{\hbar} \sum_{mn}  \int_0^{t} d\tau \bar \psi_{m}(\tau) \left[ i \hbar \partial_t\delta_{mn} - f^0_{m n} \right]\psi_{n}(\tau)}}  \nn\\
&\equiv&\gamma^+_{ab}~ G_0^{f}(\tau,n|\tau',m) - \gamma^-_{ab}~G^{b}_0(\tau,n|\tau',m)\, ,
\ee
 where  $G_0^{f} (n, \tau|m\tau') $ and $ G_0^{b} (n, \tau|m,\tau') $  are called the forward and backward time-directed propagators and  obey the equations
\be
 (i \hbar \partial_t -f_{nm}) G_0^{f/b} (n, \tau|m,\tau') = \mp i \hbar \delta(\tau-\tau') ~\delta_{nm} \, .
\ee
As usual, to  explicitly compute the time ordered $G^0_{ab}(n,\tau|m,\tau') $ we    introduce  different Feynman's ``$i\epsilon$" prescriptions for forward and backward propagating excitons:
\be
G^{f}_0(n,\tau|m,\tau') &=& \int \frac{d \omega}{2 \pi} e^{-i \omega (\tau-\tau')} ~\frac{ -i }{\omega - \frac{ f^0_{nm}}{\hbar} + i 0^+}\\
G^{b}_0(n,\tau|m,\tau') &=& \int \frac{d \omega}{2 \pi} e^{-i \omega (\tau-\tau')} ~\frac{  i }{\omega - \frac{ f^0_{nm}}{\hbar} - i 0^+}
\ee
Here,  $V_{ij}$ is the unitary matrix that  diagonalizes $f_{mn}^0$ and $E_k$ are the corresponding eigenvalues. 
The final result for the time-ordered exciton propagator is then:
\be
G^0_{ab}(n,\tau|m,\tau')= \sum_k ~V^\dagger_{n k}~e^{-\frac{i}{\hbar}E_k(\tau-\tau')} V_{k m}~ \left(\gamma^+_{ab}~ \theta(\tau-\tau')- \gamma^-_{ab}~\theta(\tau'-\tau) \right) \, .
\ee
Finally, each vertex comes with a factor $i~\sqrt{\frac{M \gamma}{\beta}} f_{nm}^j$.  

The explicit evaluation of these leading order diagrams   gives:
\be\label{P0result}
 \rho(t)_{nm} \simeq \rho^{(0)}_{nm} (t)+ \rho^{(1)}_{nm}(t)
 \ee
 where $\rho^{(0)}_{nm}(t)$ is the unperturbed result:
 \be
\rho^{(0)}_{nm} (t) = G^0(n_0 0| n,t) ~G^0(m, t| m_0, 0),
 \ee
 while $\rho^{(1)}_{nm}(t)$ is the leading order correction and reads
 \be\label{P1corr}
 \rho^{(1)}_{nm}(t) & = & - \frac{2 M \gamma}{\beta \hbar^2}~ \sum_{s q s' q' } \sum_{ij} \int_0^t ~ d\tau \int_0^t ~ d\tau'  \left\{ \right.
 G^{0 }(n_0,0| s', \tau') ~f^{j}_{ s' q'} ~G^{0}(q',\tau'|n,0) ~\Delta_{j i} \LL( \tau'-\tau \RR) ~G^{0}(m,t|s,\tau)~f^{i}_{ sq} ~G^{0}(q,\tau| m_0,0)\nn \\
 &+& G^{0}(m, t| m_0,0)~G^{0}(n_0,0| q', \tau)~f^{j}_{ q' s'} ~\Delta_{ji} (\tau-\tau')     ~G^{0}(s' ,\tau|s, \tau')~f^{i}_{ s q} ~G^{0}(q, \tau'| n,t) \nn \\
&+& G^{0}(n_0, 0| n,t) G^{0}(m, t| q', \tau) ~f^{j}_{ q' s'} ~\Delta_{ji} (\tau'-\tau)     ~G^{0}(s' ,\tau'|s, \tau)~f^{i}_{ s q} ~G^{0}(q, \tau| m_0,0)\left.\right\}.
\ee
The first line is the ``one-vibron" exchange contribution, while the following two lines are the  ``self-energy" diagrams.

An important comment concerns the normalization of the density matrix. 
In our previous work  \cite{Schneider1}, we adopted the perturbative normalization  scheme, where we first computed the perturbative correction to the free normalization, i.e.   $\text{Tr}\hat \rho\simeq 1+ x$, and then we used the approximation $\frac{1}{\text{Tr}\rho}\simeq \frac{1}{\text{Tr}\rho^0 + \text{Tr} \rho^{(1)}}=\frac{1}{1+x}\simeq 1-x$.  Clearly, in this scheme,  the density matrix is only  approximately normalized. 
Instead, in this work we choose to adopt an exact normalization, i.e. we divide the un-normalized matrix computed in perturbation theory by its trace. We have found that, in this alternate scheme, the range of time intervals in which a perturbative estimate of $\rho_{nm}(t)$ is reliable is improved.

We conclude this section by emphasising that, in this perturbative approach, it is straightforward to study the contribution to the density matrix which follows from the interaction of excitons with  specific vibronic normal modes. Indeed, to this goal it is sufficient to retain only some terms of the sum over normal mode frequencies entering the expression for the vibron propagator (\ref{Delta}).

\section{Non-perturbative calculation of the density matrix}\label{non-perturbative}
In the presence of strong exciton-vibron couplings and in the long-time regime, the perturbative approach described in the previous section becomes inapplicable. 
Thus, in this section, we develop an alternative and rigorous nonperturbative approach. To this goal,  it is convenient to return to the expression Eq. (\ref{generating}) and carry out the Gaussian path integral over the exciton fields  $\psi$ and $\bar \psi$ . This leads to an equivalent  expression for the generating functional in terms of the vibron field only:
\be\label{Zeff}
Z[\eta, \bar \eta]= \int \mathcal{D}\delta r~e^{-\left(S_{OM}[\delta r] +\text{Log Det}G^{-1}_{\delta r}\right)}~e^{\frac{i}{\hbar}\sum_{mn}\int_0^t d\tau~ \bar \eta(n;\tau) ~G_{\delta r}(n,m;\tau;\tau')~ \eta( m;  \tau')}\, ~e^{-\beta H_{MM}(0)}.
\ee
where $G_{\delta r}(n,m;\tau, \tau') $ is the  time-ordered propagator for the electronic excitations in the background of the vibration field, i.e.
 \be\label{Gdeltar}
G_{\delta r}(n,\tau|m,\tau')&=& 
 \frac{ \int \mathcal{D}\bar \psi \mathcal{D} \psi~\psi^a_{n}(\tau)~\bar \psi^b_{m}(\tau')~ 
  e^{\frac{i}{\hbar} \sum_{mn}  \int_T^{T+t} d\tau \bar \psi_{m}(\tau) \left[ i \hbar \partial_t\delta_{mn} - f^0_{m n} - \sum_i \delta r_i f^i\right]\psi_{n}(\tau)} }
 { \int \mathcal{D}\bar \psi \mathcal{D} \psi~ e^{\frac{i}{\hbar} \sum_{mn}  \int_T^{T+t} d\tau \bar \psi_{m}(\tau) \left[ i \hbar \partial_t\delta_{mn} - f^0_{m n} - \sum_i \delta r_i f^i \right]\psi_{n}(\tau)}} \nn\\
&\equiv&\gamma^+_{ab}~ G_{\delta r}^{f}(\tau,n|\tau',m) - \gamma^-_{ab}~G^{b}_{\delta r}(\tau,n|\tau',m)\, ,
\ee
where the forward and backward propagators $ G^{f/b}_{\delta r}$ obey the equations
\be\label{EqGf/b}
 (i \hbar \partial_\tau - f^0_{nm} - \sum_i f^{i}_{nm} \delta r_i(\tau)) )~G^{f/b}_{\delta r}(\tau,n|\tau',m) = \mp~i~ \hbar ~\delta(\tau-\tau') \delta_{nm} \, .
\ee

We emphasise that in Eq.(\ref{Zeff}),  the classical Langevin dynamics of the nuclei ---which is controlled by the $S_{OM}$ functional---  is modified by the  back-action exerted by the electronic excitations on the molecule's conformation, which is controlled by the functional
 \be
 S_{back}[\delta r] \equiv \text{Log Det}G^{-1}_{\delta r}.
 \ee
where
\be
G^{-1}_{\delta r} = \frac{i}{\hbar} (i \hbar \partial_t - f_{nm}^0 -  f_{nm}^i \delta r^i(\tau) )\, .
\ee

The  explicit representation of the back-action term is derived in the appendix A. Expanding in powers of the fluctuation field $\delta r$ and truncating to quadratic order (as is done for the classical OM action) we obtain 
\be\label{Sback}
\log[ \det G^{-1}_{\delta r_i}]
&\simeq& \text{const.}+\frac{1 }{2} \sum_{ij}\sum_{s\ne t}~C^i_{st} C^j_{ts}~\int_0^t d\tau \int_0^td\tau' \delta r^i(\tau)~\delta r^j(\tau')~\cos\left[(\tau-\tau')~\frac{E_t-E_s}{\hbar}\right]\nn\\
\ee
In this expression, 
$
C^k_{s t} \equiv \frac{1}{\hbar} \sum_{m n} V_{s m} f^k_{m n } V_{n t}^\dagger
$
are coefficients which express the coupling of different vibronic modes to the electronic transitions. 
Therefore, our final result is:
\be
\log[ \det G^{-1}_{\delta r_i}] &=& \text{const.} +\frac{1 }{2} \sum_{ij}\sum_{s\ne t}~C^i_{st} C^j_{ts}~\int_{0}^{t} d\tau \int_{0}^{t} d\tau' \delta r^i(\tau)~\delta r^j(\tau')~\cos\left[ 2 \pi ~\frac{(\tau-\tau')}{t}~(N_s-N_t)\right].\nn\\
\ee
It is instructive to consider the corresponding expression in Fourier space:
\be
\log[ \det G^{-1}_{\delta r_i}] &\simeq& \text{const.} +  \frac{1}{2} \sum_{s \ne t} \sum_{ij}~C^i_{t s}~C^j_{s t}\int \frac{d \omega}{2\pi}~ b^i_\omega~b^j_{-\omega} ~2\pi~ \delta\left(\omega-\frac{E_{s}-E_{t}}{\hbar}\right)
\ee
where $b^i_\omega$ is the Fourier transform of the vibration field $\delta r_i$. In this form it is evident that only vibrations with the correct energy can couple to quantum electronic transitions. Since  the time integration  is  bounded between $0$ and $t$,  the energy difference $(E_{s}-E_{t})/\hbar$ must be rounded in order to match a multiple of the elementary frequency unit $\Delta \omega = 2 \pi/t$, which sets the energy resolution power which can be achieved letting the system evolve for a time $t$. In other words 
\be
\frac{(E_s- E_t)}{\hbar} \rightarrow 2 \pi \frac{N_s-N_t}{t}, 
\ee
where $N_s =  \min_{\{ k\}} \left| \frac{E_s}{\hbar} - \frac{ 2\pi}{t}~k~\right|$.

Our final explicit representation of the generating functional defined in Eq. (\ref{Zeff}) is
\be\label{ZFour}
 Z[\bar \eta, \eta] &=& 
 \int \mathcal{D}\delta r^i~ e^{-S_{OM}[ \delta r]}~e^{-S_{back}[ \delta r]}~e^{-\beta H_{MM}(0)}~e^{\frac{i}{ \hbar}\sum_{nm}\int_{0}^{t} d\tau \int d \tau'  \bar \eta_n G_{\delta r }(n, \tau| m, \tau') \eta_m(\tau')},
\ee
where $S_{OM}$ and $S_{back}$ are defined in Eq. (\ref{SOM}) and Eq. (\ref{Sback}) respectively and  $G_{\delta r }$ is defined by Eq.s (\ref{Gdeltar}) and (\ref{EqGf/b}).

From this generating functional it is immediate to obtain the expression  for the reduced density matrix for an arbitrary number $n_E$ of quantum electronic excitations, by functionally differentiating $4 n_E$ times with respect to the sources and then setting them to 0. Our final result is: 
  \be\label{paved}
\rho_{n_1 \ldots  n_{n_E} m_1, \ldots, m_{n_E}}(t)  = \frac{\int \mathcal{D}\delta r^i~ e^{-S_{OM}[ \delta r] }~ e^{-S_{back}[ \delta r] }~e^{-\beta H_{MM}(0)}
 \prod_{k=1}^{n_E}~\left( G^f_{\delta r}(n_k, t | n_k^0,0)~ G^b_{\delta r}(m^{0}_k, 0|  m_k,t) \right)}{\int \mathcal{D}\delta r^i~ e^{-S_{OM}[ \delta r] }~e^{-S_{back}[ \delta r] }
 ~e^{-\beta H_{MM}(0)}}\, .
\ee 
 
 \subsection{Stochastic estimate of the density matrix}\label{stochALG}

The expression (\ref{paved})  paves the way for defining an algorithm which provides a stochastic estimate of the density matrix at any finite time $t$:
\begin{enumerate}
\item An ensemble of $N$ thermally equilibrated  initial conditions $\{\delta r^{(k)}_i(0)\}_{k=1\ldots,N}$ is generated either by directly sampling the Boltzmann distribution $e^{-\frac{\beta}{2} \sum_{ij} \delta r_i(0) \mathcal{H}_{ij} \delta r_j(0)} $,  or  by running $N$ long classical Langevin simulations. 

\item These $N$ initial conditions are used to generate $N$ independent Markov chains of molecular trajectories, according to the following rules.  Trial moves in the Markov chain  are proposed by  integrating the classical Langevin equation defined by the OM functional starting from the same initial condition.
Hence, each trial move corresponds to an entirely new Langevin trajectory $\delta r^i_{new}(\tau)$ with the same initial condition $\delta r^i(0)$ and lasting for a time interval $t$. Trial moves are accepted or rejected by comparing the back-action terms calculated using the old and the new Langevin trajectory, i.e. using the standard Metropolis condition:
\be
\eta <\min\left[1, e^{-S_{back}[\delta r^{(k)}_{new}] +S_{back}[\delta r^{(k)}_{old}]}\right],
\ee
where $\eta\in [0,1]$ is a random number sampled from a uniform distribution. After equilibration, this procedure  yields $N$ statistically independent trajectories, $\{\delta r^{(k)}_i(\tau)\}_{k=1,\ldots, N}$. 

\item  For each of the $N$ equilibrated paths  $\{\delta r^{(k)}_i(\tau)\}_{k=1,\ldots, N}$, the forward and backward propagators $G^f_{\delta r_i^{(k)}}$ and $G^b_{\delta r^{(k)}_i}$  are calculated by solving the one-dimensional time-dependent  Schr\"odinger equation forward and backward in time (see Eq.   (\ref{EqGf/b})). A discussion of the numerical calculation of  $G^f_{\delta r_i^{(k)}}$ and $G^b_{\delta r^{(k)}_i}$ is given in appendix \ref{appendixB}.
Consequently, $N$ independent realizations of the  reduced density matrix are obtained from
\be\label{rhorel}
\rho^{(k)}_{n_1 \ldots  n_{n_E} m_1, \ldots, m_{n_E}}(t)   =  \prod_{k=1}^{n_E}~\left( G^f_{\delta r^{(k)}_i}(n_k, t | n_k^0,0)~ G^b_{\delta r^{(k)}_i}(m^{0}_k, 0|  m_k,t) \right).
\ee
 
\item 
Finally, the stochastic estimate of the density matrix is computed in the standard way, by relying on the Central Limit theorem, i.e. by averaging over the $N$  independent   realizations (\ref{rhorel}):
\be
\rho_{n_1 \ldots  n_{n_E} m_1, \ldots, m_{n_E}}(t)  \simeq \frac{1}{N} \sum_{k=1}^N \rho^{(k)}_{n_1 \ldots  n_{n_E} m_1, \ldots, m_{n_E}}(t)  
\ee  
The statistical error is estimated from the variance:
\be
\delta \rho_{n_1 \ldots  n_{n_E} m_1, \ldots, m_{n_E}}(t) \simeq \frac{1}{\sqrt{N-1}} \left( \frac{1}{N} \sum_{k=1}^N( \rho^{(k)}_{n_1 \ldots  n_{n_E} m_1, \ldots, m_{n_E}}(t) )^2 -
(\frac{1}{N} \sum_k \rho^{(k)}_{n_1 \ldots  n_{n_E} m_1, \ldots, m_{n_E}}(t)  )^2\right)^{1/2}
\ee
\end{enumerate}
We emphasise that  this algorithm does not suffer from any dynamical sign or phase problem.

For realistic values of masses, couplings and temperatures we observed that the back-action effective functional is often found to be very small,  so that the exponential factor $e^{-S_{back}}$ undergoes only small fluctuations around $1$. In this cases the relatively expensive Monte Carlo procedure described above can be avoided. Instead, it is sufficient to  directly  compute the un- normalized density matrix from an average over the 
product of exciton propagators with the exponential factor $e^{-S_{back}}$, i.e. 
\be\label{reweight}
\rho_{n_1 \ldots  n_{n_E} m_1, \ldots, m_{n_E}}(t)  \simeq \textrm{const.}\cdot \frac{1}{N} \sum_{k=1}^N~\left[ \rho^{(k)}_{n_1 \ldots  n_{n_E} m_1, \ldots, m_{n_E}}(t) ~e^{-S_{back}^{(k)}}\right]
\ee
Then, the normalization constant must be determined \emph{ a posteriori}, from  the trace of the density matrix.

We conclude this section by stressing that physical insight may be gained by analysing the effect  of specific vibrational normal modes on the propagation of the electronic excitations.
For example, this type of analysis can be useful to test if the propagation of electronic excitations in some photosynthetic complex  is supported by the so-called noise-assisted 
quantum transport mechanism ~\cite{Plenio_Nat}. 

Also in this nonperturbative approach it is possible to single out the effect on the excitonic dynamics due the coupling to specific normal modes, by replacing the Hessian $\mathcal{H}_{ij}$ matrix, which  enters into the definition of the $S_{OM}$ functional and of the Boltzmann weight, with a modified Hessian matrix $\overline{\mathcal{H}}_{ij}$,  constructed by adding up only the contributions of the selected set of normal modes:
\be\label{projectout}
\mathcal{H}_{ij}  \rightarrow \overline{\mathcal{H}}_{ij} \equiv \sum_{k = N_i}^{N_s}~U^\dagger_{i k}~ M \Omega_k^2~ U_{k j}
\ee
The summation on the right-hand side runs over the selected set of normal modes,  $\Omega_{N_i}, \Omega_{N_i+1}, \ldots,  \Omega_{N_s}$.

\section{Quantum dissipative dynamics at the mesoscopic scale}\label{ETtheory}
So far we have described the dynamics of the macromolecule  using the model Hamiltonian (\ref{Htot}), which is characterized by a  space resolution of the order of the size of the molecular fragments, $\sigma_{\text{short}} \sim$ nm. 
Let us  now discuss the case in which the macromolecular system  extends into the mesocopic length-scale regime,  i.e. has size  $S \sim \mu$m. Notable examples of such systems include organic semi-conductors made of inter-digited conjugate polymers.   In this section  we discuss how using RG one can build a much more coarse-r-grained ET, which is guaranteed to provide the \emph{same asymptotic IR dynamics} of the model defined by Eq. (\ref{Htot}), yet is far simpler and can be applied in the mesoscopic regime.  As usual, this new ET contains  effective parameters which have to be determined from   the microscopic theory (\ref{Htot}), or directly fitted against experimental data. This matching involves choosing a so-called renormalization point where both microscopic and ET are expected to hold. 

The derivation of the mesoscopic ET starting from  our original model Hamiltonian (\ref{Htot})  is quite elaborate, and was carried out in detail Ref. \cite{Schneider2}. Here, we only review  the underlying assumptions  and summarize the results. We consider the case in which the macromolecular system is probed at a very low resolution scale   $\lambda\gg \sigma_{\text{short}}$, i.e.  much larger than the length scale at which its chemical structure   is revealed. We also imagine to study the dynamics at times  much longer than those at which quantum coherence is washed out. 
More specifically, let us consider the ratio between the excitation's  DeBroglie's wavelength $\lambda_B$ and the resolution scale $\lambda$:
 \be
 \xi = \frac{\lambda_B}{\lambda}
 \ee
 At low resolution power $\xi\ll1$,  one can specify the position of the excitation using a continuum vector ${\bf x}$, rather than the discrete molecular orbital  index. Furthermore,   it is sufficient to specify the diagonal elements of the density matrix, since quantum coherence is lost.  

In Ref. \cite{Schneider2}, it is rigorously shown that to leading order in an expansion in  $\xi\ll1$, the path integral yielding the conditional probability $P({\bf y}, t| {\bf x}, t_0)$ to observe the excitation at ${\bf y}$ at time $t$ provided it was observed at ${\bf x}$ at $t=0$, reduces to that of a simple diffusion process:
\be\label{ETPI1}
P({\bf y}, t| {\bf x}, 0) = \int_{\bf x}^{\bf y} \mathcal{D} {\bf R}~ e^{-\int_0^td\tau\sum_{ij} 
\dot{R}_i \frac{1}{4}D_{ij}^{-1}\dot{R}_j}\ee
where $D^{-1}_{ij} \equiv g_{ij}~\bar{D}^{-1}$ is an effective diffusion tensor to be determined by the renormalization.
However, keeping into account of effects of order $\xi^2$, the path integral takes the following form 
\be
\label{ETPI2}
P({\bf y}, t| {\bf x}, 0) = \int_{\bf x}^{\bf y} \mathcal{D} {\bf R}~ e^{-\int_0^td\tau\sum_{ij} 
\left[\dot{ R}_i \frac{1}{4}D_{ij}^{-1}\dot{R}_j +  \dot{ R}_i^2 C_{ij} \dot{R}_j^2 \right]}
\ee
Where $C_{ij} \equiv  g_{ij} ~\bar C$ introduces one additional effective parameter, $\bar C$. We note that, in these expressions, the explicit dependence on the expansion parameter $\xi^2$ has been  absorbed in the definition of the effective tensors $D^{-1}_{ij}$ and $C_{ij}$,  which have to be  independently determined  (see \cite{Schneider2} for more details). The additional term, proportional to $\dot R^4$ encodes some quantum effects which are not captured by Eq. (\ref{ETPI1}).

Let us now discuss the renormalization of this ET, i.e. the matching between microscopic at mesoscopic theories, to order $\xi^2$ accuracy. For sake of simplicity, here we consider the case of a one-dimensional system (molecular wire). The generalization to multidimensional case is straightforward and is discussed in the original publication. 
With the stochastic algorithm discussed in  section \ref{stochALG}, one can microscopically compute  the first two moments of the probability distribution $P(m,t)$ of finding the excitation at some fractional orbital $m$, at time $t$ i.e. 
\be
 M_2(t) &\equiv& \sum_{m}  a^2~\text{Tr}[|m\ra \la m| \hat \rho(t)]~(m- m_0)^2 \\
 M_4(t) &\equiv& \sum_{m}  a^4~\text{Tr}[|m\ra \la m| \hat \rho(t)]~(m- m_0)^4,
\ee
where $\hat \rho(0)  =  |m_0\ra \la m_0| \times e^{-\beta H_{MM} + H_{CL}}$ is the initial density matrix ---cf. Eq. (\ref{imtime})--- and $a$ is the distance between consecutive molecular orbitals in the molecular wire.  These predictions must be set equal to that of the mesoscopic ET (\ref{ETPI1}):
\be
M_2(t)=\la \Delta R^2(t) \ra &=& 2 \bar D~t\\
M_4(t)=\la \Delta R^2(t) \ra &=& 60 \bar D^2 t^2 - \bar C t.
\ee
Clearly, the matching must be imposed at some time scale $t_R$, where both the microscopic and the mesoscopic theories are expected to hold, namely when the dependence of $M_2(t)$ on $t$ is found to be  linear.   Note that,  to lowest-order accuracy in the $\xi^2$ expansion,  one only needs to evaluate $M_2(t_R)$ and extract the effective diffusion constant $\bar D$. 

In passing, we note that it is possible to obtain a simple analytic expression for the conditional probability (\ref{ETPI2}) to any order of accuracy in $\xi^2$. To order $\xi^2$ one finds
\be\label{solution}
P(x, t| x_i) \simeq \sqrt{\frac{1}{ 4 \pi D t}} ~e^{-\frac{(R-R_i)^2}{4 D t}}~\left[1- C \left(\frac{(x-x_i)^4}{t^3 D}-20\frac{(x-x_i)^2}{t^2} + 60 \frac{1}{t}\right) \right]
\ee
 
 The mesoscopic ET provides an example of rigorous multi-scale theory, in which the microscopic physics is systematically encoded in the effective parameters and the structure of the operator reflects the symmetries of the underlying (more) fundamental physics. 
\section{Application: Exciton propagation on a linear molecular chain}\label{application1}

Before considering an application to a realistic system, it is useful to  assess the accuracy of our perturbative and nonperturbative field-theoretic techniques on a benchmark case. To this end, we consider the model for exciton propagation on a linear molecular chain  which has been introduced and solved numerically within a semiclassical approximation by Iubini and co-workers~\cite{Piazza}.
The Hamiltonian of this system consists of the sum of three contributions:
\be
\hat H = \hat H_e + \hat H_L + \hat H_s.
\ee
where
\be
\hat H_e = \sum_{n=1}^L~ e_n(u)~ \hat a^\dagger_n \hat a_n + \sum_{n=1}^{L-1}~J_n(u)~(\hat a^\dagger_{n+1} \hat a_n + \hat a^\dagger_{n} \hat a_{n+1})
\ee
 is the tight-binding Hamiltonian of an exciton propagating on a one-dimensional  lattice, whose configuration is defined by the set of coordinates $ u=(u_1, \ldots, u_L)$.  Each beed  in this lattice represents a monomer in a linear polymer. The coupling between excitons and monomers is obtained by linearising the coefficients $J_n(u)$ and $e_n(u)$:
 \be
e_n(u) &\simeq& E_0 + \chi_E~(u_{n+1}-u_{n})\\
J_n(u) &\simeq& J_0 + \chi_J~(u_{n+1}-u_{n})
 \ee
 $\hat H_L$ is  the Hamiltonian which controls the vibrations of the chain and read
 \be
 \hat H_L= \sum_{n=1}^{L-1} \left[\frac{\hat p^2_n}{2M} + \frac{\kappa}{2}\left(\hat u_{n+1}-\hat u_n\right)^2\right]
 \ee
Finally, $\hat H_b$  is a  Caldeira--Leggett Hamiltonian for an Ohmic bath which represents the solvent.  In practice, its specific form is irrelevant. Indeed, once the classical limit on the chain dynamics is taken, the role of the heat bath  Hamiltonian is only to turn the Newton's equation for the motion of the residues into a Langevin equation with viscosity constant $\gamma$. 
The numerical values of the parameters of this model are reported in table I. 

 \begin{table}[t!]
\begin{center}
\begin{tabular}{|c| c c c c c c c c|}
\hline
parameter: & $M $ & $\gamma $ & $T $ & $ \kappa $ & $E_0$ & $J_0$& $\chi_E$ & $\chi_J $ \\
\hline
value: & 1  & 1 &  1 & 1  & 1 &  1  & 1 & 0 \\
\hline
  \end{tabular}
  \caption{The parameters of the model for a linear molecular chain of $L=20$ residues. All values are given in appropriate 
  powers of eV, assuming  natural units in which $\hbar=k_b=c=1$.} 
\end{center}
\label{default}
\end{table}%

\subsection{Benchmarking the non-perturbative calculation against the semiclassical solution}

Let us first use this model to  assess the accuracy of our nonperturbative approach against the results obtained by Iubini and co-workers in Ref. \cite{Piazza}. In their work, the authors adopted a semiclassical scheme based on coupling the  Newton's equation for the  coordinates $u_n(t)$ with a Schr\"odinger equation for the exciton's wave function. 
They used the Langevin dynamics  to  thermalize the chain and then switched off  noise and viscosity, letting the chain evolve according to the Newton's equation, i.e. in the gas phase. 
By contrast, in our approach, the monomers experience the effect of fluctuations and dissipations not only during the initial  thermalization, but also  throughout the entire time evolution of the system. In addition to the classical harmonic forces, the monomer dynamics is also influenced by exciton back-action term.   We integrated the Langevin  using a elementary timestep $\Delta t=0.02$~fs and the solved the forward and  backward  Schr\"odinger equations using the algorithm presented in appendix B, diagonalising the Hamiltonian every $0.02$~fs. The average was performed over 1000 independent Langevin trajectories, and the effect of the back-action term was accounted for by re-weighting, as in  Eq.~(\ref{reweight}).   We found that the effect of back-action was to  generate small to the  statistical weight of the Langevin trajectories, $0.95 \lesssim e^{-S^{(k)}_{back}[u(\tau)]} \lesssim 0.98 $.

\begin{figure}[t!]
\includegraphics[width=8.5cm]{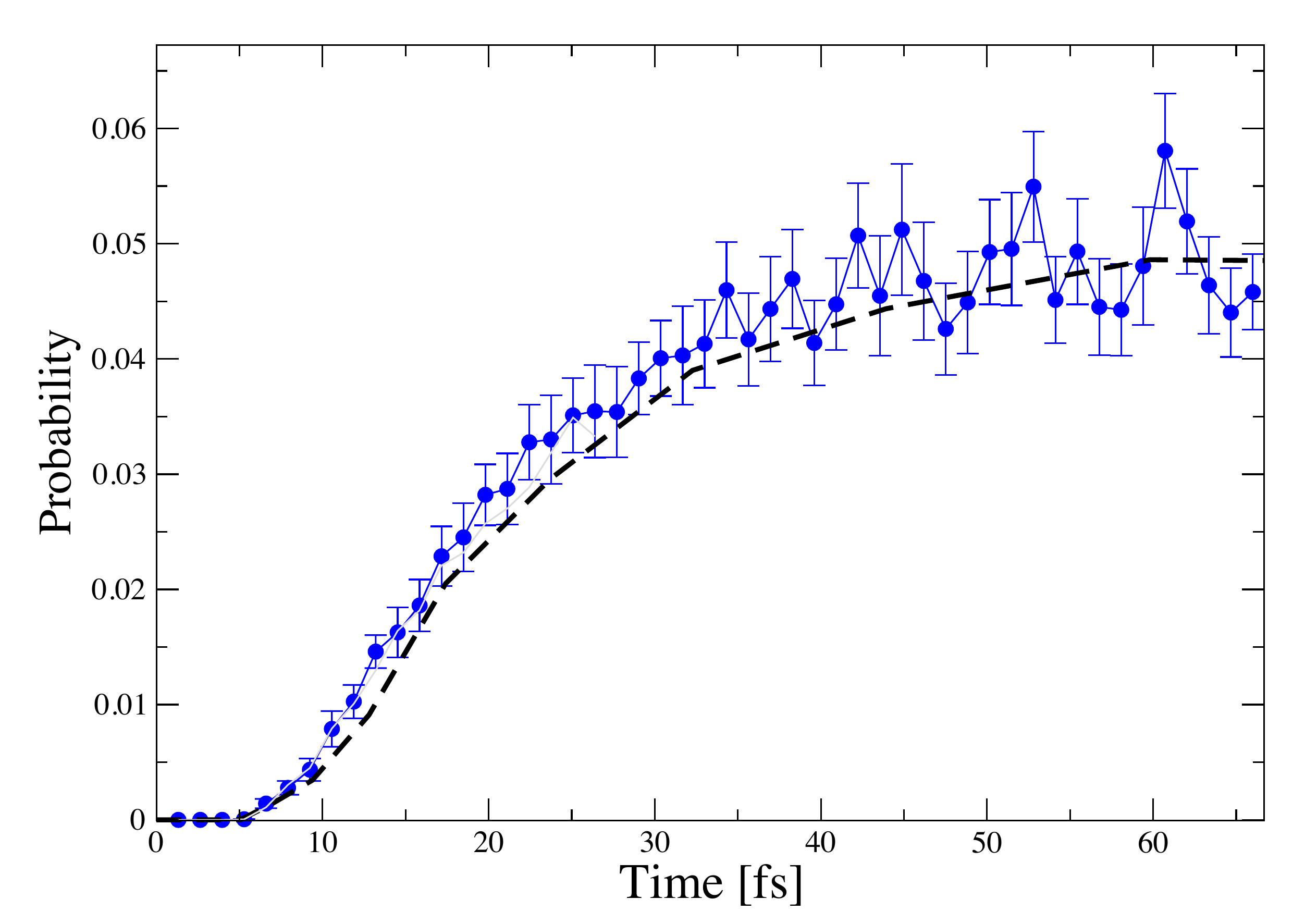}
\includegraphics[width=8.5cm]{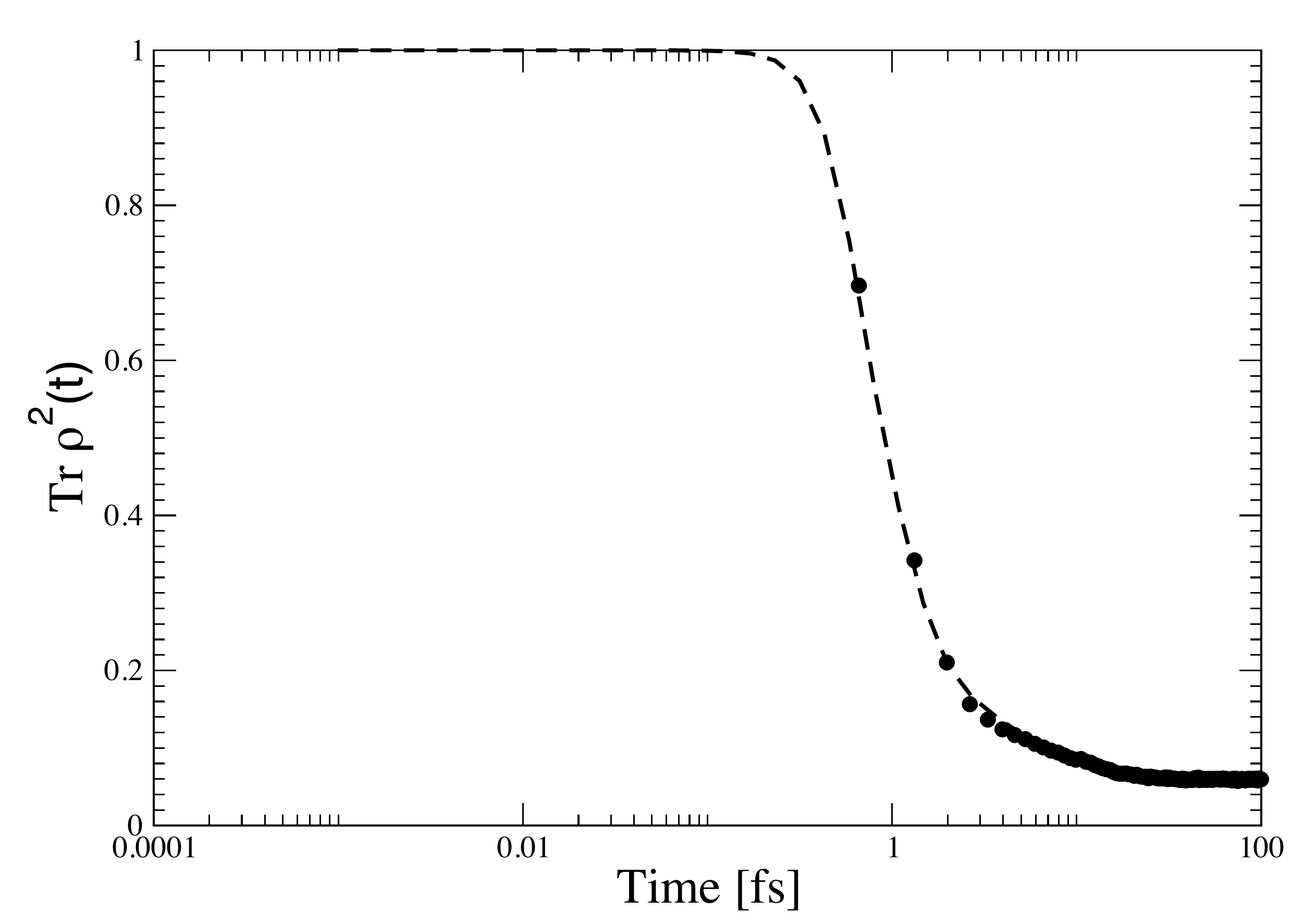}
\caption{Left panel: Probability that a hole injected at the leftmost  monomer of the chain at time 0 is observed at the rightmost  monomer at time $t$. Right panel: : Time evolution of $\text{Tr}\rho^2(t)$. The points are the result of our non-perturbative stochastic approach while the dashed line is the result of the semiclassical calculation described in Ref.~\cite{Piazza}.}
\label{P20}
\end{figure}

In the left panel of Fig. \ref{P20} we compare the probability that an exciton injected in the leftmost monomer is observed in the opposite endpoint of the chain after a time $t$, calculated in our method and in the semiclassical approach of Iubini \emph{et al.}. The results are consistent, within statistical errors. In particular, both methods predict that  thermalization is achieved after about  $60$~fs.

In the right panel, we compare the $\textrm{Tr}[\rho^2(t)]$ which is often used as a measure of the loss of quantum coherence. In this case, the agreement between the two calculations is excellent, indicating that  quantum coherence is lost after a few fs. 

\subsection{Comparing perturbative and nonperturbative calculations}

\begin{figure}[t!]
\includegraphics[width=10cm]{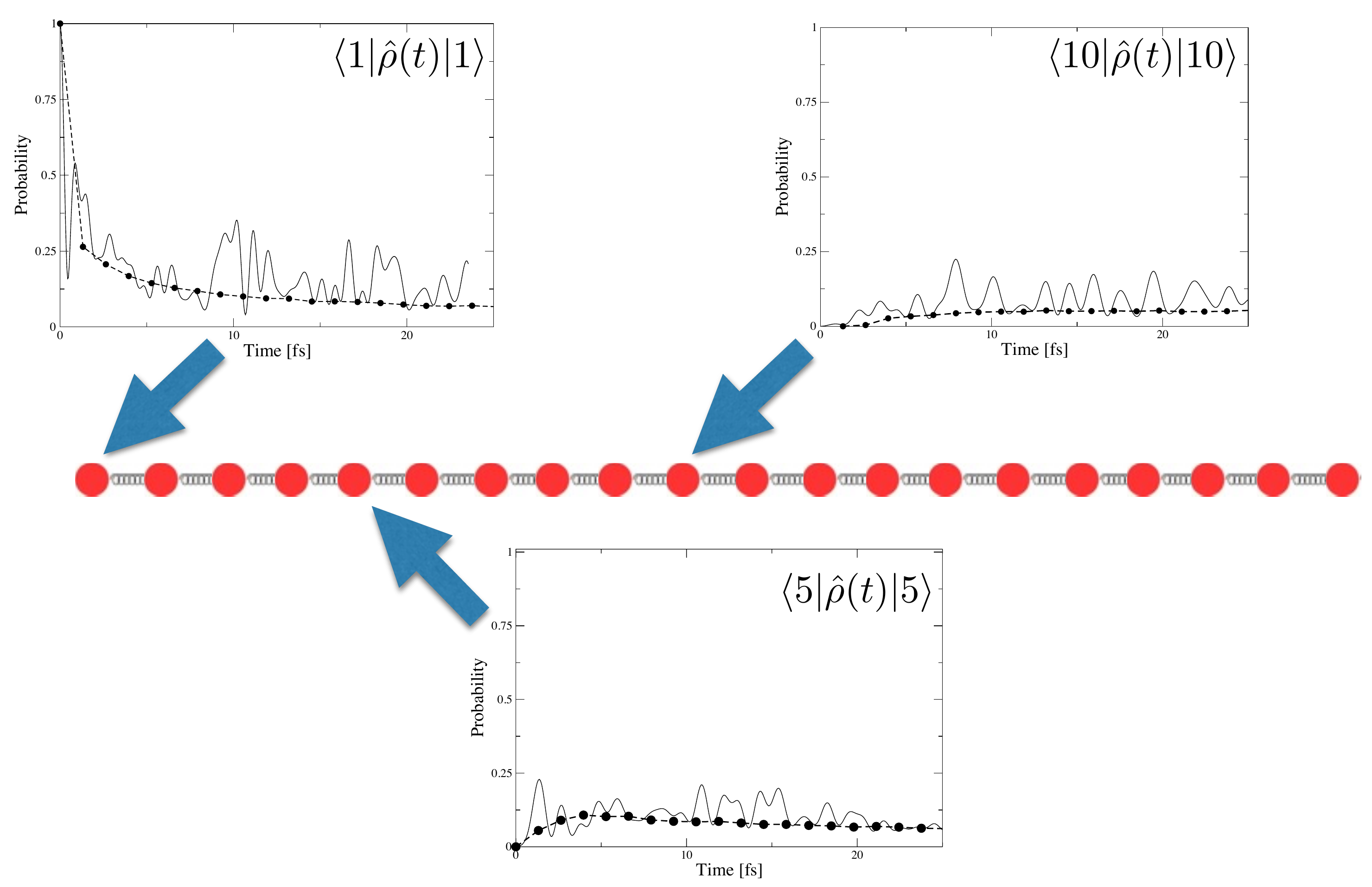}
\caption{ Probability that a hole injected at the leftmost  monomer of the chain at time 0 is observed in 3 different monomer at time $t$. The points are the result our non-perturbative stochastic approach while the solid line are the estimates obtained to leading order in  perturbation theory. }
\label{PT}
\end{figure}

Having validated  our nonperturbative approach against an alternative numerical method, we are in a condition to study the accuracy of the simple leading order perturbative estimate, obtained in a  semi--analytic way, i.e. from the Feynman diagrams calculated in section \ref{perturbative}.
In Fig. \ref{PT} we show  the comparison between the perturbative and non-perturbative probabilities of observing the exciton in different locations of the chain as a function of time $t$. As in the previous calculation, the exciton was injected in the leftmost residue at time $t=0$. 

These  results show that the perturbative calculation captures the overall  behavior of the different probabilities at the semi-quantitative levels. Most interestingly, a single vibron exchange is sufficient to force the probabilities to oscillate around their equilibrium values. For longer times, however, oscillations become much wider,  signaling the breakdown of the perturbative approximation. 

The corresponding predictions for the evolution of $\text{Tr}\rho^2(t)$ are compared in the right panel. Again we see that the perturbative estimate is able to predict the loss of quantum coherence induced by the vibronic correlations. Direct inspection of the contribution of the different diagrams shows that the decoherence is driven by the one-vibron exchange diagram, i.e. by the first perturbative diagram shown in Fig. \ref{two point}.

\section{Application: Multi-scale Calculation of Hole Mobility in  a Conjugated Polymer}\label{application2}

Let us now apply the theory developed in this work to a realistic physical system. Our main goal is to illustrate in practice how the multi-scale approach discussed in the previous sections  makes it possible to consistently encode in a single physical model the physics emerging at different  length scales, ranging from the  \AA~to the $\mu$m.

\begin{figure}[t!]
\includegraphics[width=16cm]{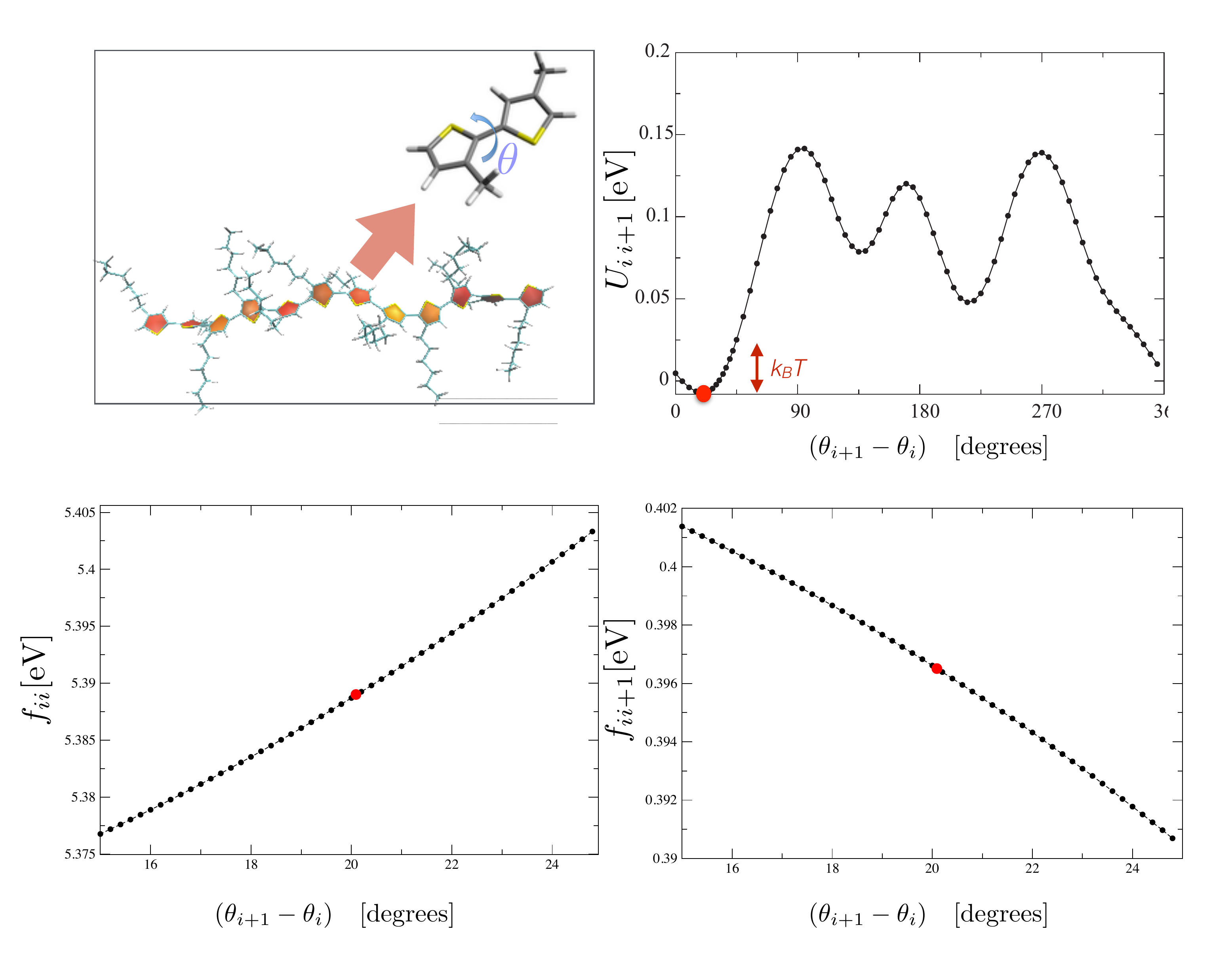}
\caption{Upper-left corner: structure of the 8-residue P3HT polymer. Upper right corner: Pairwise molecular potential energy as a function of the relative ring plane angle between  neighboring residues, obtained from DFT-B calculations. Lower-left (-right) panel: On-site energy (transfer integral) as a function of the relative ring angle between  neighboring residues, obtained from DFT-TB calculations. The red spots identify the configuration of minimum molecular potential energy. }
\label{P3HT}
\end{figure}

 To this end, we  consider intra-chain propagation of electron holes through the backbone of a poly-3-hexylthiophene (P3HT) chain (see upper-left panel of Fig.~\ref{P3HT}).  Among the different families of conjugated polymers which can be used to process and study organic semiconductors, P3HT polymers are  particularly attractive in view of their relatively large charge carrier mobility, their relatively facile processing and good environmental stability.

Our goal is to study the propagation over long P3HT molecular wires in solutions, and compute the hole's mobility.
To this end we need to consider a molecular wire which is sufficiently long for the quantum dissipative dynamics to reach the anomalous diffusion regime given by Eq. (\ref{ETPI2}), before  electronic excitations injected in the middle of the chain reach either of the chain terminals. To ensure this condition, we studied a  polymer consisting of $150$ residues and we assumed that a single electronic excitation (hole) is injected in the midpoint  at time $t=0$.  

We adopted a microscopic theory in the form of Eq. (\ref{Htot}) in which the molecular fragments are identified with the monomers. The total molecular potential energy (first Born--Oppenheimer energy surface) is assumed to depend only on the dihedral angles $ (\theta_1, \ldots, \theta_L)\equiv \Theta$ which specify  the orientation of the plane of the  aromatic rings  (see upper-right panel of Fig.~\ref{P3HT}).  This assumption is justified by the fact that the side-chains tend to align perpendicularly to  the backbone,  thereby setting no significant steric constraints. The total molecular energy is approximated with the sum of pairwise terms, each depending on the  relative  angle between consecutive aromatic rings, $(\theta_{i+1}-\theta_i)$:
\be
V(\Theta) = \sum_{i=1}^{L-1} U_{i i+1}(\theta_{i+1}-\theta_i)
\ee
The pairwise interaction energy function $U_{i i+1}(\theta_{i+1}-\theta_i)$ was obtained from electronic structure calculations, i.e. solving the Schr\"odinger equation in  the Born--Oppenheimer approximation  using the  Density Functional Theory Tight Binding (DFT-TB) formalism \cite{frauenheim2000self, aradi2007dftb+}. The results are shown in the upper-right corner of Fig.~\ref{P3HT}. We found that, in the lowest--energy configuration,  neighbouring aromatic rings   form a relative dihedral angle of $(-1)^i\theta_0$, where $\theta_0\simeq 20^o$ and $i$  is the monomer index.
 
In a small-angle expansion around this equilibrium configuration, the Hamiltonian describing conformational vibrations takes the form
\be\label{HM}
 H_M \equiv  \sum_{i=1}\frac{p_i^2}{2 I} + \sum_{i=1} \frac{\kappa}{2}\left[\theta_{i+1}-\theta_i + (-1)^i \theta_0 \right]^2 \ee
In this equation, $p_i$ is the conjugate momentum  and $I$ is the moment of inertia of a monomer, which can be computed directly from the chemical structure and is found to be $I=3.4\times 10^3$~amu~\AA$^2$. The spring constant $\kappa$ was extracted from  an harmonic fit of the mechanical potential energy $U_{i i+1}(\theta_{i+1}-\theta_i)$, near the lowest energy configuration and found to be $\kappa =0.3$~eV rad$^{-2}$. The last two terms in Eq.~(\ref{HM}) come from assuming that the first and last monomer in the chain are bonded to some external nonconducting leads which align them horizontally.

To describe the propagation of charge carriers, in our case holes, we assigned a fractional molecular  orbital to each of the aromatic rings of the  poly-3-hexylthiophene molecule. Our choice is motivated by the fact that $\pi$-electrons are highly delocalized on the aromatic ring. We then computed the on-site and transfer integrals in the generalized Mulliken-Hush theory  \cite{dedachi2007charge}.  We  took the value of the highest occupied molecular orbital (HOMO) energy of the thiophene dimer as the on-site energy, and half the difference between the highest occupied molecular orbital (HOMO) energy and the next-to-HOMO lower molecular orbital (HOMO-1) energy of the dimer as the transfer integral:
\be
  f_{nm} &=& \frac{1}{2} ( E_{\text{HOMO}} - E_{\text{HOMO}-1} )( \delta_{nm}-1) + E_{\text{HOMO}} \delta_{nm}  
\ee
To obtain the first derivative of the on-site energies and of the transfer integrals as a function of the dihedral angle $\theta$, we simply calculated their values for a series of molecular configurations of the dimer, corresponding to a grid of values for $\theta$, and then fitted the resulting table.
The results are shown in the lower-left and lower-right panels of Fig.~\ref{P3HT}. From these calculations we extracted the coefficients $f_{nm}(\Theta_0)$ which define the Frenkel Hamiltonian $H_0$:
\be
\hat H_0 \equiv \sum_{nm}~f_{nm}(\Theta_0)~ \hat a^\dagger_n \hat a_m
\ee
and the  coefficients $ f^i_{nm} \equiv \frac{\partial}{\partial \theta_i} f_{nm}(\Theta_0)$ which define the strength of the coupling of the holes to the vibrations.  In particular, we have used the following  parameterization:
\be \begin{cases}
f_{nm}^0  = J_0~(\delta_{n m+1}+\delta_{n m-1}) - E_0~ \delta_{n m}\\ 
f^k_{nm} =  T_1~ (\delta_{n m+1} + \delta_{n m-1}) ~\delta_{k n} + T_0~ \delta_{n m}~\delta_{k n}
\end{cases}
\ee
where 
$J_0=0.4$~eV, $E_0=5.4$~eV, $T_0=0.15$~eV rad$^{-1}$ and $T_1= 0.06$ eV rad$^{-1}$ .  The inter-residue distance $a$ is 0.4 nm. 
 
The viscosity parameter $\gamma$ depends on the specific molecular  environment in which the polymer is embedded and can be evaluated by computing the velocity auto-correlation function by molecular dynamics simulations. In  this analysis we arbitrarily chose  $\gamma =0.1$~ps$^{-1}$ which is of  the order of magnitude expected for a polymer chain in solution water.

The vibronic-electronic interaction in this system is expected to be very different from  the molecular chain studied in the previous example. Indeed, in the conjugated polymer the normal mode  spectrum and the spectrum of electronic energy differences are  separated by a gap of several orders of magnitude~(see  the inset in the left panel of  Fig. \ref{P3HT_prob}). 
In the left panel Fig.~\ref{P3HT_prob} we show the  probability $P_{\textrm{end}}(t)$ that a hole  injected in the center of the chain  (monomer 75)  at time $t=0$ is detected  at the leftmost terminal (monomer 1) at time $t$. This plot shows that the time it takes the hole to reach the end of the chain is about $55-70$~fs. Thus, any calculation of the mobility must be performed in a regime of times smaller than $60$~fs to avoid finite--size effects. 
In the right-hand side of Fig.~\ref{P3HT_prob} we report the time evolution of  Tr$[\rho^2]$. We see that, in this system, quantum coherence is significantly damped after about 40 fs. 
Hence, the matching between microscopic theory and mesoscopic ET must be carried out at a renormalization time $t_R$ in the range  $40~\textrm{fs}~\lesssim t_R \lesssim~55~$fs. 

\begin{figure}[t!]
\includegraphics[width=8 cm]{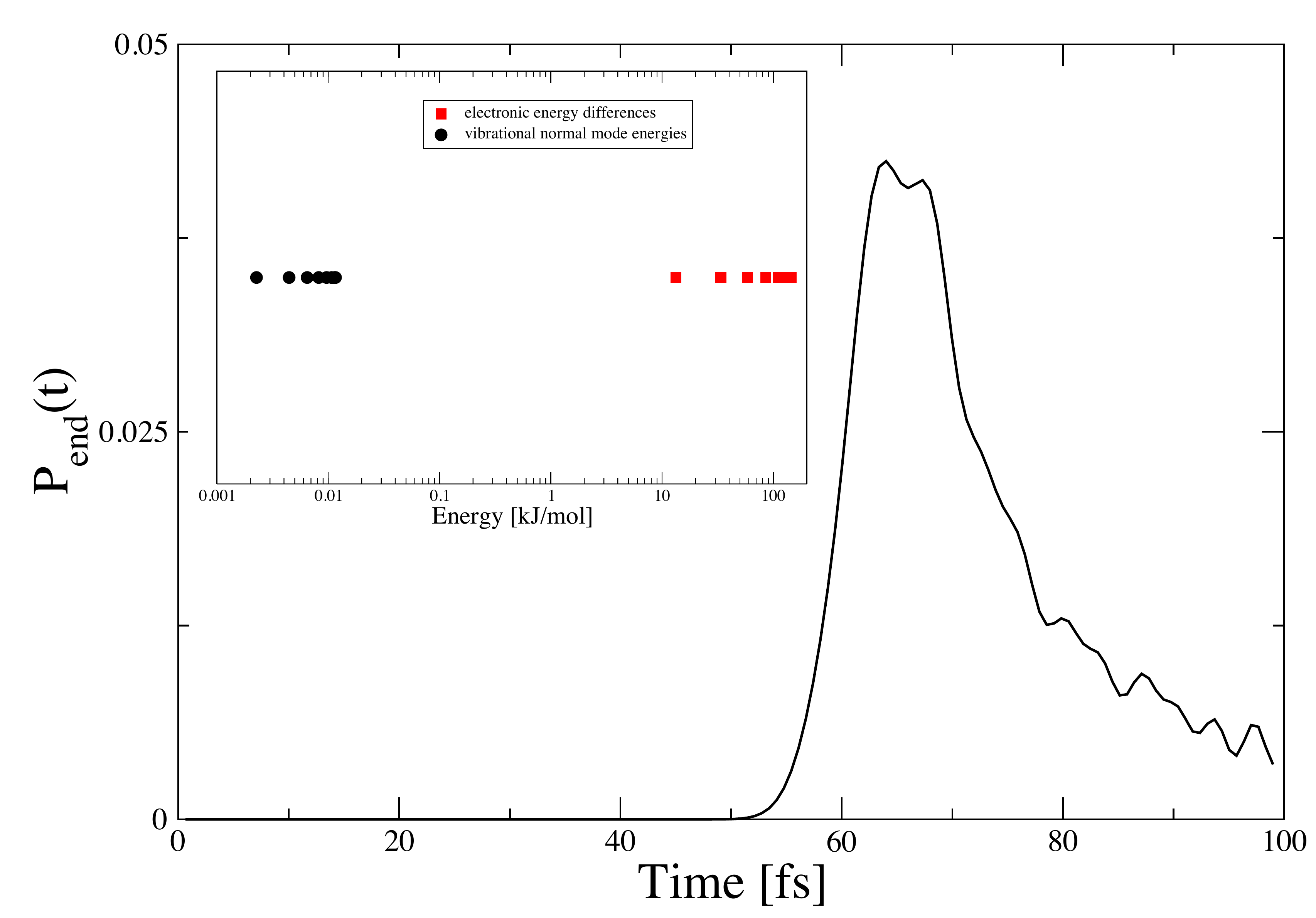}\quad
\includegraphics[width=8cm]{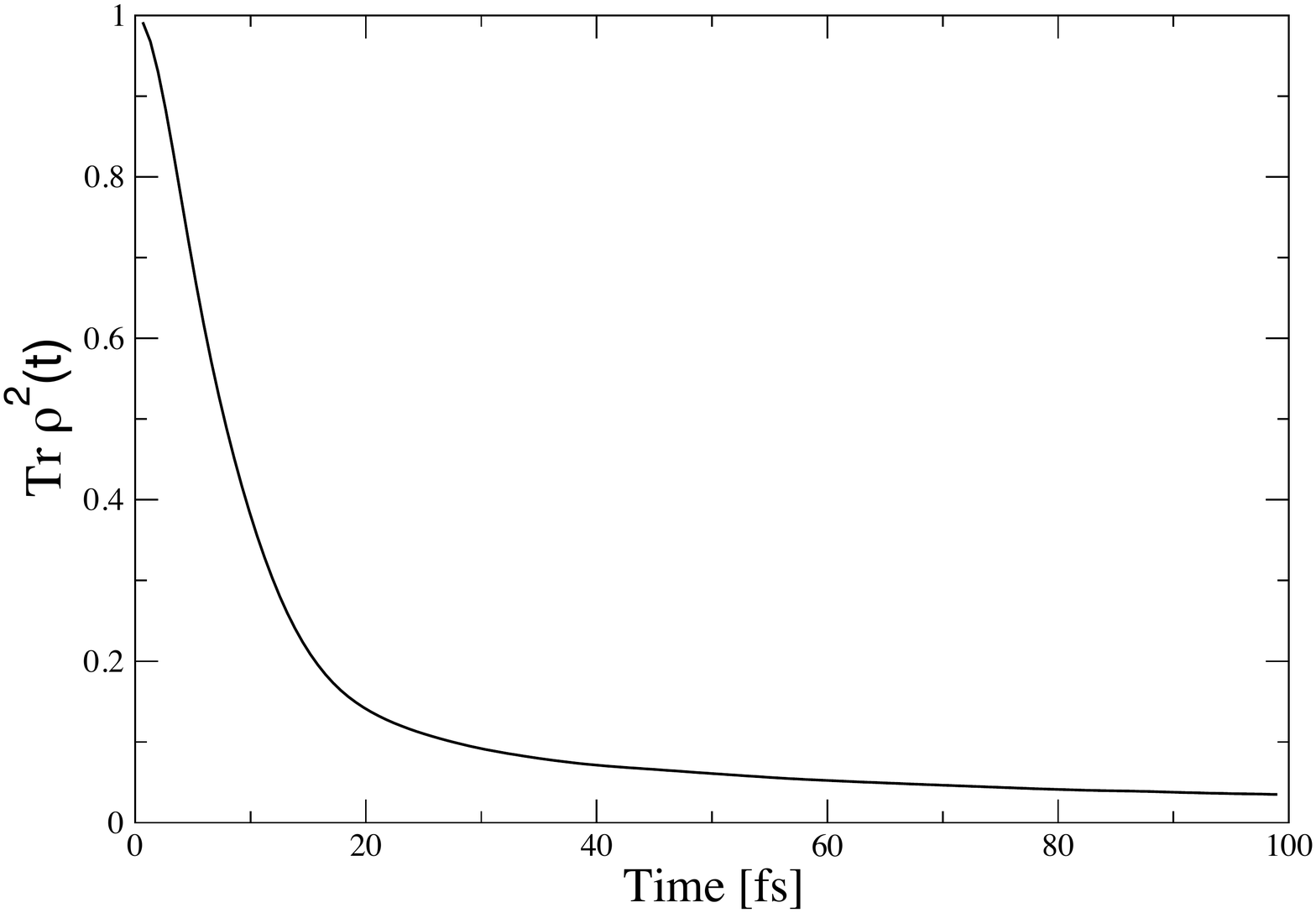}
\caption{Left panel: Probability that a hole injected at the centre of a 150 residue P3HT polymer (residue index 75) at time $0$ is observed at leftmost end of the chain (residue index 1) at time $t$. The inset at the upper left corner compares the energy of the vibronic normal modes with the electronic energy differences. Right panel: Time evolution of $\text{Tr}\rho^2(t)$, quantifying the loss of quantum coherence.  }
\label{match}
\end{figure}

In Fig. \ref{match} we see that in this range the scaling of $M_2(t)$ with $t$ is indeed approximatively linear, while in $M_4(t)$ non-linear effects are more pronounced, as expected. 
Implementing the matching at  $t_R=45~$fs, we obtain
\be
\bar D \simeq 11 \textrm{nm}^2~\textrm{ps}^{-1} \qquad \bar C \simeq -1330~  \textrm{nm}^4 \textrm{ps}^{-1}
\ee 
We note that also in a previous  application to homo-DNA wires, the two coefficients were found to have opposite signs. 

The effective diffusion parameter can be used to estimate the hole's mobility  $\mu$, using the  linear-response theory relation
\be
\mu = \frac{e}{k_BT} \bar{D} 
\ee
 For $T=300K$ we obtain the estimate $\mu \simeq 4~\textrm{cm}^2 \textrm{s}^{-1} \textrm{V}^{-1}$, which is a value of the same magnitude of typical mobility measurements in organic seminconductor~\cite{Podzorov,Bredas_rev} and in very good agreement with mobility predicted with different theoretical models~\cite{Troisi_2006,Troisi_transition}. 

\section{Conclusions}
\label{conclusions}
We have developed a systematic multi-scale approach to investigate the non-equilibrium dissipative quantum  dynamics of electronic excitations which propagate in macromolecular systems in solution. We have introduced and compared different approaches, which can be used to obtain predictions at different levels of space and time resolution. 
In particular, the perturbative scheme and  the Path integral Monte Carlo approach can be used to perform quantum transport calculations in realistic macromolecular systems while adopting a microscopic model in which all parameters and coupling are derived bottom-up from the underlying electronic structure. The perturbative approach is, of course, by far the least computationally expensive, at least to leading order. It also offer explicit analytical insight on the dependence of the computed density matrix on the physical parameters, such as temperature, viscosity, etc \ldots The Monte Carlo algorithm does not suffer  from any dynamical sign or phase problem, thus  it is quite computationally efficient and can be used to study realistic systems for time intervals as large as ps.  Finally, we showed how the Monte Carlo calculation can be interfaced with a rigorous low-resolution representation of the dynamics obtained using RG argument. The result is an ET which can be used to study quantum transport in the mesoscopic regime. 
  
\begin{figure}[t!]
\caption{$M_2(t)$ (left panel) and $M_4(t)$ (right panel) for the 150 residue P3HT polymer, computed microscopically using our Monte Carlo algorithm.  The red dot is the point used to match with the mesoscopic ET ( corresponding to the renormalization timescale $t_R=45$fs). }
\label{P3HT_prob}
\end{figure}

To conclude, let us provide some details on the computational cost of performing our Path Integral Monte Carlo simulations: Completing the calculation on the polymers considered in this work required minutes on a  standard laptop computer. A fortran code implementing this algorithm can be made available upon request to the corresponding author.

\acknowledgments
We thank F. Piazza and S. Iubini for sharing their numerical results for the time evolution of the density matrix used in section \ref{application1}. PF acknowledges special support from University of Trento through the grant ``Bando Progetti Strategici di Ateneo". 

{}
\newpage
\appendix

\section{Calculation of  $\log \det$ to Order $\delta r^2$. }\label{appendixA}
In this appendix we provide the details of the explicit calculation of the  quantum back--action term $\log[ \det G^{-1}_{\delta r_i}]$  up to quadratic order in $\delta r$.  It is convenient to introduce the Fourier transform of the vibrons: 
\be\label{F}
b^i_{\omega} &=& \int  d\tau ~e^{i \omega \tau}~\delta r^i(\tau)\\
\delta r^i(\tau) &=&  \int \frac{d \omega}{2\pi} e^{-i \omega \tau} b^i_\omega,
\ee
where for convenience we have identified  the Fourier series with its continuous integral limit.

The zero-th order contribution is dynamically irrelevant and can be ignored. The term  of order $\delta r$ vanishes identically:
\be
\log[ \det G^{-1}_{b_i}] ^{-1}&\simeq&  \text{const.}+\sum_i \int d\tau \frac{\delta}{\delta r_i(\tau) }\left.\log\det\left[\frac{i}{\hbar}(i\hbar \partial_t- f_{mn}-f^i_{mn} \delta r_i)\right]^{- \frac{1}{2}}\right|_{\delta r=0} \delta r^i(\tau)\nn\\
&=& \text{const.} -i \sum_i \int d\tau \sum_{nm} \frac{f^{i}_{n m}}{\hbar}~\delta r_i(\tau) G_0(\tau, n; \tau, m)\nn\\
&=& \text{const.} - t \sum_i  \sum_{nm} \frac{f^{i}_{n m}}{\hbar}~b_{0}^i~\int \frac{d \omega'}{2\pi}\left( \frac{1}{\omega' -\frac{ f_{nm} }{\hbar}  + i0^+} + \frac{1}{\omega' - \frac{f_{nm}}{\hbar} - i0^+}\right)=0\nn\\
&=& \text{const.}-  \lim_{\epsilon\to 0} \sum_i  \sum_{nm}  ~f^{i}_{n m} ~b_{0}^i~\sum_s V_{n s} V^\dagger_{ s m}\left(\int_{E_s+\epsilon}^\infty\frac{d \omega'}{2\pi} +\int^{E_s-\epsilon}_{-\infty} \frac{d \omega'}{2\pi}\right) \frac{t}{ \hbar \omega' -E_s } =0
\ee

The  leading contribution comes at quadratic order in $\delta r$:
\be\label{TRLOG1}
\log[ \det G^{-1}_{b_i}]^{-1}&\simeq& \text{const.}+\frac{i^2}{2 \hbar^2} \sum_{ij}\sum_{nmn'm'} \int d\tau' ~\int d\tau \delta r_i(\tau) \delta r_j (\tau') ~f^i_{n m}~f^j_{m' n'}
\Delta^{(2)}(n,m; \tau|n',m', \tau')\\
\ee
Where $\Delta^{(2)}(n,m; \tau|n',m', \tau')$ is the two-point correlation function in the free theory and reads
\be\label{Delta2}
 \Delta^{(2)}(n,m;\tau|n',m';\tau')&\equiv&\sum_{b a}G^{ba}_0(n',\tau'|n, \tau) G^{ab}_0(m, \tau|m', \tau')\nn\\
  &=& \sum_{ab} ( G^f_0(n',\tau'|n,\tau) \gamma^+_{ab}- G^b_0(n',\tau'|n,\tau) \gamma^-_{ab}~)~(G^f_0(m,\tau|m',\tau') \gamma^+_{ba}- G^b_0(m,\tau|m',\tau') \gamma^-_{ba}~ )\\
  &=&\sum_{nmn'm'} ~\int \frac{d w}{2\pi}\int \frac{d w'}{2\pi}~e^{i (\tau'-\tau)(w' -w)} \left(
\frac{-i \hbar}{\hbar w-f_{n'n}+i0^+}\frac{-i \hbar}{\hbar w'-f_{m' m}+i0^+}+\frac{i \hbar}{\hbar w-f_{n'n}-i0^+}\frac{i \hbar}{\hbar w'-f_{m' m}-i0^+}\right) \, , \nn\\
\ee
We find
\be\label{TRLOG2}
\log[ \det G^{-1}_{b_i}]^{-1}&\simeq& \text{const.}+\frac{-1}{2 \hbar^2} \sum_{ij}\sum_{nmn'm'}\int \frac{dw'}{2\pi} \int \frac{dw}{2\pi}~C^i_{st} C^j_{ts} b^i_{w'-w} b_{w-w'}^j\nn\\
&&~\left(\frac{-i \hbar}{\hbar w-E_s+i0^+}\frac{-i \hbar}{\hbar w'-E_t+i0^+}+\frac{i \hbar}{\hbar w-E_{s}-i0^+}\frac{i \hbar}{\hbar w'-E_t-i0^+}\right) \nn\\
&=& \text{const.}+\frac{-1}{2} \sum_{ij}\sum_{st}\int \frac{d\omega}{2\pi} \int \frac{dw}{2\pi}~C^i_{st} C^j_{ts}~ b^j_{\omega}~ b_{-\omega}^i\nn\\
&&~\left(\frac{-i \hbar}{\hbar w-E_s+i0^+}\frac{-i \hbar}{\hbar (w-\omega)-E_t+i0^+}+\frac{i \hbar}{\hbar w-E_{s}-i0^+}\frac{i \hbar}{\hbar (w-\omega)-E_t-i0^+}\right) 
\ee
In this equation we have  introduced
some coefficients $C^k_{a b}$ which express the coupling of the $k-$th vibron to the electronic transition between the  energy levels $a$ and $b$,
\be
C^k_{s t} \equiv \frac{1}{\hbar} \sum_{m n} V_{s m} f^k_{m n } V_{n t}^\dagger.
\ee
Using Cauchy's integral formula, we obtain
\be\label{TRLOG3}
\log[ \det G^{-1}_{b_i}]^{-1}&=& \text{const.}-\frac{1}{2} \sum_{ij}\sum_{st}\int \frac{d\omega}{2\pi} ~C^i_{st} C^j_{ts}~ b^j_{\omega}~ b_{-\omega}^i~\left(~\frac{i }{\omega-(E_t-E_s)/\hbar+i0^+} + \frac{-i}{\omega-(E_t-E_s)/\hbar-i0^+}\right) \nn\\
&=& \text{const.}-\frac{1}{2} \sum_{ij}\sum_{st}\int d\omega~C^i_{st} C^j_{ts}~ b^j_{\omega}~ b_{-\omega}^i~\delta\left[\omega-\frac{(E_t-E_s)}{\hbar}\right]\\
&=& \text{const.}-\frac{1 }{2} \sum_{ij}\sum_{st}~C^i_{st} C^j_{ts}~b^j_{\frac{E_t-E_s}{\hbar}}~ b_{-\frac{E_t-E_s}{\hbar}}^i\\ \ee
Finally, transforming back to the time representation:
\be\label{TRLOG4}
\log[ \det G^{-1}_{b_i}]^{-1}&=& \text{const.}-\frac{1 }{2} \sum_{ij}\sum_{st}~C^i_{st} C^j_{ts}~\int d\tau \int d\tau' \delta r^i(\tau)~ \delta r^j(\tau')~ e^{i (\tau-\tau') \frac{E_t-E_s}{\hbar}}\nn\\
&=& \text{const.}-\frac{1 }{2} \sum_{ij}\sum_{st}~C^i_{st} C^j_{ts}~\int d\tau \int d\tau' \delta r^i(\tau)~\delta r^j(\tau')~\cos\left((\tau-\tau')~\frac{E_t-E_s}{\hbar}\right)\nn\\
\ee

The terms in this sum with $s=t$ should be remove, because they do not involve any energy exchange between the vibrons and the exciton and introduce an infrared divergece.  We now show that  such a divergence is associated with the diffusion of the center of mass. To this end, let us consider the contribution to the back-action coming from the terms with $i=j$ and $s=t$:
\be
\lim_{\omega\to 0}~\frac{( C^i_{ss})^2}{2}~\sum_i b^i_\omega b^i_{-\omega}
\ee
At mean-field level, the summation $\sum_i b^i_\omega b^i_{-\omega}$ is approximated by the spectral density $S_{ii} \equiv \langle  b^i_\omega b^i_{-\omega}\rangle$, where the average is evaluated over different realization of the Langevin noise. 
After transforming into  the phonon basis, in which the Hessian matrix is diagonal,  spectral density reads:
\be
S_i(\omega) \equiv \la  b^i_\omega b^i_{-\omega}\ra =  \frac{2 k_BT \gamma}{M} \sum_{i,k}  U_{n i}  \frac{1}{(\omega^2-\Omega_k^2)^2 + \gamma^2 \omega^2} U^\dagger_{k i}  
\ee
where $\Omega_k$ are the normal mode frequencies. For $\omega \to 0$ we find
\be
\lim_{\omega\to 0} S_i(\omega) =  \frac{2 k_BT \gamma}{m}~ \sum_{k}  U_{i k}  \frac{1}{\Omega_k^4}U^\dagger_{k i}
\ee
This quantity diverges because of the zero-mode $\Omega_0=0$,  associated with the centre of mass motion.  

Finally,  we emphasise that when restoring the finite intervals in the time integration $\int d\tau \to \int_{0}^{t} d\tau$ we need to approximate the energy difference $(E_{s}-E_{t})/\hbar$, in order to match multiples of the elementary frequency mode (resolution power) $\Delta \omega = 2 \pi/t$. In other words 
\be
\frac{(E_s- E_t)}{\hbar} \simeq (N_s-N_t)  \frac{2\pi}{ t}, 
\ee
where $N_s =  \min_{\{ k\}} \left| \frac{E_s}{\hbar} - \frac{ 2\pi}{t}~k~\right|$.

\section{Numerical Evaluation of the Propagators.}\label{appendixB}
The algorithm developed in this work requires to solve the time-dependent Schr\"odinger equation forward and backward in time in the background of the 
same molecular vibration field $\delta r_k(\tau)$. Here, we provide some details on how this can be implemented, in practice. 
 
Suppose we are interested in computing the density matrix element $\rho_{nm}$ at time $t$, assuming the excitation is initially injected at the fractional orbital $m_0$ and let $\delta r_k(\tau)$ be a trajectory of molecular vibrations, obtained by integrating the Langevin equation. 

We split the time interval $t$ into a large number  $N_t$ of very short time intervals. The number of time intervals (thus their size  $\Delta t= t/N_t$)  must be chosen in such a way  that $\delta r_i(\tau)$ can considered  approximately static for $\tau\in [t, t+\Delta t]$. In practice, $\Delta t$ can be chosen to coincide with the time step used to integrate the Langevin equation. 

Starting from the initial condition $|m_0\rangle \langle m_0|$, we construct the forward and backward propagators by applying  $N_t$ times the corresponding elementary evolution operators: 
\be\label{time_evol}
G^f_{\delta r}(n,t|m_0,0) &\equiv&\langle n| \prod_{i=0}^{N_t-1} e^{-\frac{i\Delta t}{\hbar}  H[\delta r_k(t_{i})] \Delta t} |m_0\rangle,\\
G^b_{\delta r}(m_0,0|m, t) &\equiv&\langle m| \prod_{i=0}^{N_t-1} e^{+\frac{i\Delta t}{\hbar} H[\delta r_k(t_{i})] \Delta t} |m_0\rangle.
\ee 
In these equations, the matrix $ H[\delta r(t_i)]$ is the electronic  Hamiltonian represented in the fractional orbital basis, computed in the background of the  external vibration $\delta r_k(\tau)$ evaluated at time $  t_i \equiv i \cdot \Delta t$:
\be
 H_{nm}[\delta r_k(t_i)] = f^0_{nm}+ \sum_k f^k_{nm} \delta r_k(t_i).
\ee  
We stress that, since the intrinsic time scales characterising the electronic dynamics are usually much shorter than those characterising the  vibrations, it is convenient to evaluate the elementary time evolutions in Eq. (\ref{time_evol}) by explicitly  diagonalising $H_{lm}[\delta r_k(t_{i})]$, and avoid linearising the evolution operator. 

The matrix element $\rho_{nm}(t)$ is then simply the product of the forward and backward propagators:
\be
\rho_{nm}(t) =G^f_{\delta r}(n,t|m_0) \cdot G^b_{\delta r}(m_0,0|m,t). 
\ee

\end{document}